\documentclass[11pt]{article}

\usepackage{bigstrut}
\usepackage{fullpage}
\usepackage{amsfonts,color,amssymb,graphicx,stmaryrd}
\usepackage{amsmath}
\usepackage{tikz}
\usetikzlibrary{arrows,backgrounds,decorations,decorations.pathmorphing,positioning,fit,automata,shapes,matrix,patterns}
\usepackage{booktabs}
\usepackage[htt]{hyphenat}
\usepackage{amsmath, amssymb, amsfonts}
\usepackage{color,graphicx, enumerate, subfig}
\usepackage{mathtools}
\usepackage{array}
\usepackage{scalefnt}
\usepackage{bm}
\usepackage{xspace}
\usepackage{stmaryrd}
\usepackage{upgreek}
\usepackage{multirow}
\usepackage{mathrsfs}
\usepackage{url}
\usepackage{bbm}
\usepackage{algorithmic,algorithm}
\usepackage{eufrak}
\usepackage{bigstrut}
\newtheorem{theorem}{Theorem}
\newtheorem{lemma}[theorem]{Lemma}

\newtheorem{example}[theorem]{Example}

\def\Proof{{\bf Proof.}}
\def\qed{{\bf $\Box$}}
\newcommand{\mypar}[1]{\subsection{#1}}

\usepackage{tikz}
\usetikzlibrary{arrows}
\usetikzlibrary{decorations.pathmorphing}
\usetikzlibrary{backgrounds}
\usetikzlibrary{positioning}
\usetikzlibrary{fit}
\usetikzlibrary{petri}

\tikzstyle{background}=[fill=gray!20, inner sep=0.2cm]

\newcommand{\Fig}[1]{Fig. #1}

\def\bigO{{O}}
\def\qed{{\bf $\Box$}}
\newcommand{\st}{\ensuremath{^*}}

\newcommand{\domain}{\ensuremath{\mathbb{D}}}

\newcommand{\posrat}{\ensuremath{\mathbb{Q}^+}}

\newcommand{\Nat}{\ensuremath{\mathbb{N}}}
\newcommand{\Rat}{\ensuremath{\mathbb{Q}}}
\newcommand{\Int}{\ensuremath{\mathbb{Z}}}

\newcommand{\ignore}[1]{}

\def\hole{?}
\newcommand{\vocab}{F}

\newcommand{\trees}{T}
\newcommand{\ptrees}{T^{\hole}}

\newcommand{\CG}{G}
\newcommand{\CF}{{\mathbb F}}
\newcommand{\CCF}{{\mathbb F}^c}
\newcommand{\sep}{\ensuremath{\,|\,}}

\newcommand{\reg}[1]{{\mathbb R}(#1)}

\newcommand{\treesof}[1]{T_{#1}}
\newcommand{\CostModel}{\ensuremath{\mathbb{C}}}

\newcommand{\sradd}{\oplus}
\newcommand{\srmul}{\otimes}

\newcommand{\SST}{{\sc\textsc sst}}

\newcommand{\SSTT}{{\sc\textsc sstt}\xspace}

\newcommand{\stt}{\ensuremath{U}}

\newcommand{\inputalph}{\ensuremath{\Sigma}}

\newcommand{\state}{\ensuremath{q}}

\newcommand{\fm}[1]{\ensuremath{#1^{*}}}

\newcommand{\valuation}{\ensuremath{\nu}}

\newcommand{\interp}[1]{\ensuremath{\llbracket #1\rrbracket}}

\newcommand{\computation}[1]{\ensuremath{\llbracket #1\rrbracket}}

\newcommand{\exptime}{\ensuremath{\sc\textsc{ExpTime}}\xspace}

\newcommand{\nlogspace}{\ensuremath{\sc\textsc{NLogSpace}}}
\newcommand{\ptime}{\ensuremath{\sc\textsc{PTime}}\xspace}

\newcommand{\np}{{\sc\textsc NP}\xspace}
\newcommand{\nphard}{{\sc\textsc NP}-{\sc\textsc Hard}\xspace}

\newcommand{\EDWA}{{CRA}\xspace}
\newcommand{\REDWA}{{CRA}-{RLA}\xspace}

\newcommand{\edwa}{\ensuremath{M}}

\newcommand{\edwastates}{\ensuremath{Q}}
\newcommand{\edwastate}{\ensuremath{q}}
\newcommand{\edwainitst}{\ensuremath{\edwastate_0}}
\newcommand{\edwavariables}{\ensuremath{X}}

\newcommand{\edwatrans}{\ensuremath{\delta}}
\newcommand{\edwavarup}{\ensuremath{\rho}}

\newcommand{\edwafinal}{\ensuremath{\mu}}

\newcommand{\edwavaluation}{\ensuremath{\nu}}

\newcommand{\edwaexpr}[1]{\ensuremath{E(#1)}}
\def\pexpr{E^\hole}

\newcommand{\WA}{{\sc\textsc wa}\xspace}

\newcommand{\srset}{\ensuremath{\domain}}
\newcommand{\srplus}{\ensuremath{\oplus}}

\newcommand{\srzero}{\ensuremath{\bar{0}}}
\newcommand{\srone}{\ensuremath{\bar{1}}}

\newcommand{\wa}{\ensuremath{W}}
\newcommand{\wastates}{\ensuremath{P}}
\newcommand{\wainitst}{\ensuremath{I}}
\newcommand{\wafinalst}{\ensuremath{F}}

\newcommand{\watrans}{\ensuremath{E}}
\newcommand{\wainit}{\ensuremath{\lambda}}
\newcommand{\wafinal}{\ensuremath{\rho}}
\newcommand{\wapath}{\ensuremath{\pi}}

\newcommand{\secref}[1]{Sec.~\ref{sec:#1}\xspace}

\newcommand{\thmref}[1]{Theorem~\ref{thm:#1}\xspace}

\newcommand{\setof}[1]{\{#1\}}

\newcommand{\ie}{{\em i.e.}\xspace}

\newcommand{\loris}[1]{}
\newcommand{\jyo}[1]{}
\newcommand{\mux}[1]{}
\newcommand{\yyf}[1]{}

\tikzstyle{smalltext}=[font=\fontsize{7}{7}\selectfont]
\tikzstyle{captiontext}=[font=\fontsize{9}{9}\selectfont]
\tikzstyle{state}=[draw, ellipse, minimum height=5mm,
                   minimum width=5mm, inner sep=1pt,
                   text=black, font=\fontsize{8}{8}\selectfont,
                   semithick]

\def\myplus{\otimes}
\def\mytimes{\oplus}

\def\dadd{\,\underline\mytimes\,}
\def\dscale{\,\underline\myplus\,}
\def\gdplus{\,\underline{+}\,}

\begin{document}
\thispagestyle{empty}
\begin{titlepage}

\title{Regular Functions, Cost Register Automata, and Generalized Min-Cost Problems}

\author{
Rajeev Alur \and
Loris D'Antoni \and
Jyotirmoy V. Deshmukh \and
Mukund Ragothaman \and
Yifei Yuan
}

\maketitle
\thispagestyle{empty}

\begin{abstract}
Motivated by the successful application of the theory of regular
languages to formal verification of finite-state systems, there is a
renewed interest in developing a theory of analyzable functions from
strings to numerical values that can provide a foundation for
analyzing {\em quantitative\/} properties of finite-state systems.  In
this paper, we propose a deterministic model for associating costs
with strings that is parameterized by operations of interest (such as
addition, scaling, and $\min$), a notion of {\em regularity\/} that
provides a yardstick to measure expressiveness, and study decision
problems and theoretical properties of resulting classes of cost
functions.  Our definition of regularity relies on the theory of
string-to-tree transducers, and allows associating costs with events
that are conditional upon regular properties of future events.  Our
model of {\em cost register automata\/} allows computation of regular
functions using multiple ``write-only'' registers whose values can be
combined using the allowed set of operations.  We show that classical
shortest-path algorithms as well as algorithms designed for computing
{\em discounted costs\/}, can be adopted for solving the min-cost
problems for the more general classes of functions specified in our
model.  Cost register automata with $\min$ and increment give a
deterministic model that is equivalent to {\em weighted automata\/},
an extensively studied nondeterministic model, and this connection
results in new insights and new open problems.
\end{abstract}

\end{titlepage}

\section{Introduction}
\mypar{Motivation}
The classical shortest path problem is to determine the minimum-cost
path in a finite graph whose edges are labeled with costs from a
numerical domain.  In this formulation, the cost at a given step is
determined locally, and this does not permit associating alternative
costs in a speculative manner. For example, one cannot specify that
``the cost of an event $e$ is 5, but it can be reduced to 4 provided
an event $e'$ occurs sometime later.'' Such a constraint can be
captured by the well-studied framework of {\em weighted
automata\/}~\cite{Sch61,droste_handbook_2009}.  A weighted automaton
is a {\em nondeterministic\/} finite-state automaton whose edges are
labeled with symbols in a finite alphabet $\Sigma$ and costs in a
numerical domain. Such an automaton maps a string $w$ over $\Sigma$ to
the minimum over costs of all accepting paths of the automaton over
$w$.  There is extensive literature on weighted automata with
applications to speech and image processing~\cite{MPR02}.  Motivated
by the successful application of the theory of regular languages to
formal verification of finite-state systems, there is a renewed
interest in weighted automata as a plausible foundation for analyzing
{\em quantitative\/} properties (such as power consumption) of
finite-state systems~\cite{CDH10,AKL10,almagor_what_2011}.  Weighted
automata, however, are inherently nondeterministic, and are restricted
to cost domains that support two operations with the algebraic
structure of a {\em semiring}, one operation for summing up costs
along a path (such as $+$), and one for aggregating costs of
alternative paths (such as $\min$).  Thus, weighted automata, and
other existing frameworks
(see~\cite{colcombet_regular_2010,neven_finite_2004}), do not provide
guidance on how to combine and define costs in presence of multiple
operations such as paying incremental costs, scaling by discounting
factors, and choosing minimum.  In particular, one cannot specify that
``the cost of an event $e$ is 10, but for every future occurrence of
the event $e'$, we offer a refund of 5\% to the entire cost
accumulated until $e$.'' The existing work on ``generalized shortest
paths'' considers extensions that allow costs with future discounting,
and while it presents interesting polynomial-time
algorithms~\cite{goldberg_combinatorial_1988,
oldham_combinatorial_1999}, does not attempt to identify the class of
models for which these algorithmic ideas are applicable.  This
motivates the problem we address: {\em what is a plausible definition
of regular functions from strings to cost domains, and how can such
functions be specified and effectively analyzed?\/}

\mypar{Proposed Definition of Regularity}

When should be a function from strings to a cost domain, say the set
$\Nat$ of natural numbers, be considered {\em regular\/}? Ideally, we
wish for an abstract machine-independent definition with appealing
closure properties and decidable analysis questions.  We argue that
the desired class of functions is parameterized by the operations
supported on the cost domain.  Our notion of regularity is defined
with respect to a regular set $T$ of terms specified using a grammar.
For example, the grammar $t := +(t,t)\sep c$ specifies terms that can
be built from constants using a binary operator $+$, and the grammar
$t := +(t,c) \sep *(t,d)\sep c$ specifies terms that can be built from
constants using two binary operators $+$ and $*$ in a left-linear
manner.  Given a function $g$ that maps strings to terms in $T$, and
an interpretation $\interp{.}$ for the function symbols over a domain
$\domain$, we can define a cost function $f$ that maps a string $w$ to
the value $\interp{g(w)}$.  The theory of tree transducers, developed
in the context of syntax-directed program transformations and
processing of XML documents,  suggests that the class of {\em
regular\/} string-to-term transformations has the desired trade-off
between expressiveness and analyzability as it has appealing closure
properties and multiple characterizations using transducer models as
well as Monadic-Second-Order
logic~\cite{engelfriet_macro2_1999,courcelle_graph_2002,alur_stt_2011,Ho11}.
As a result, {\em we call a cost function $f$ from strings to a cost
domain $\domain$ regular with respect to a set $T$ of terms and an
interpretation $\interp{.}$ for the function symbols, exactly when $f$
can be expressed as a composition of a regular function from strings
to $T$ and evaluation according to $\interp{.}$.}

\mypar{Machine Model: Cost Register Automata}

Having chosen a notion of regularity as a yardstick for
expressiveness, we now need a corresponding machine model that
associates costs with strings in a natural way.  Guided by our recent
work on {\em streaming
transducers\/}~\cite{alur_streaming_2011,alur_stt_2011}, we propose
the model of {\em cost register automata}: a CRA is a deterministic
machine that maps strings over an input alphabet to cost values using
a finite-state control and a finite set of cost registers.  At each
step, the machine reads an input symbol, updates its control state,
and updates its registers using a parallel assignment, where the
definition is parameterized by the set of expressions that can be used
in the assignments.  For example, a CRA with increments can use
multiple registers to compute alternative costs, perform updates of
the form $x := y+c$ at each step, and commit to the cost computed in
one of the registers at the end.  Besides studying CRAs with
operations such as increment, addition, min, and scaling by a factor,
we explore the following two variants.  First, we consider models in
which registers hold not only cost {\em values}, but (unary) {\em cost
functions\/}: we allow registers to hold pairs of values, where a pair
$(c,d)$ can represent the linear function $f(n)=c+n*d$.  Operations on
such pairs can simulate ``substitution'' in trees, and allow computing
with contexts, where parameters can be instantiated at later steps.
Second, we consider ``copyless'' models where each register can be
used at most once in the right-hand-sides of expressions updating
registers at any step.  This ``single-use-restriction'', known to be
critical in theory of regular tree
transducers~\cite{engelfriet_macro2_1999}, ensures that costs (or the
sizes of terms that capture costs) grow only linearly with the length
of the input.

\mypar{Contributions}
In Section~4, we study the class of cost functions over a domain
$\domain$ with a commutative associative function $\myplus$; in
Section~5, we study the class of cost functions over a semiring
structure with domain $\domain$ and binary operations $\mytimes$ (such
as $\min$) and $\myplus$ (such as addition); and in Section~6, we
consider different forms of discounted cost functions with scaling and
addition.  In each case, we identify the operations a CRA must use for
expressiveness equivalent to the corresponding class of regular
functions, and present algorithms for computing the min-cost value and
checking equivalence of CRAs.  We summarize some interesting insights
that emerge from our results about specific cost models.  First, our
notion of regularity implies that regular cost functions are closed
under operations such as string reversal and regular look-ahead,
leading to an appealing symmetry between past and future.  Second, the
use of multiple registers and explicit combinators allows CRAs to
compute all regular functions in a deterministic manner.  Third,
despite this added expressiveness, decision problems for CRAs are
typically analyzable.  In particular, we get algorithms for solving
min-cost problems for more general ways of specifying discounting than
known before.  Fourth, since CRAs ``construct'' costs over an infinite
domain, it suffices to use registers in a ``write-only'' mode without
any tests.  This critically distinguishes our model from the
well-studied models of {\em register machines}, and more recently,
{\em data
automata\/}~\cite{kaminski_finite_1994,neven_finite_2004,bjorklund_notions_2010}:
a data automaton accepts strings over an infinite alphabet, and the
model allows at least testing equality of data values leading to
mostly negative results regarding decidability.  Fifth, it is known
that weighted automata are not determinizable, which has sparked
extensive research~\cite{mohri_weighted_2009,
kirsten_determinization_2005}.  Our results show that in presence of
multiple registers that can be updated by explicitly applying both the
operations of the semiring, classical subset construction can be
modified to get a deterministic machine.  Finally, the class of
regular functions over the semiring turns out to be a {\em strict\/}
subset of functions definable by weighted automata due to the copyless
(or linear) restriction.  It is known that checking equivalence of
weighted automata over the tropical semiring (natural numbers with
$\min$ and addition) is
undecidable~\cite{krob_equality_1992,almagor_what_2011}. Existing
proofs critically rely on the ``copyful'' nature raising the
intriguing prospect that equivalence is decidable for regular
functions over the tropical semiring.

\section{Cost Register Automata}\label{sec:cra}
\mypar{Cost Grammars}
A ranked alphabet $\vocab$ is a set of function symbols, each of which
has a fixed arity.  The arity-0 symbols, also called constants, are
mapped to domain elements.  To allow infinite domains such as the set
$\Nat$ of natural numbers, we need a way of encoding constants as
strings over a finite set of symbols either in unary or binary, but we
suppress this detail, and assume that there are infinitely many
constant symbols.  The set $\treesof{\vocab}$ of terms over a ranked
alphabet is defined in the standard fashion: if $c$ is a constant
symbol in $\vocab$, then $c \in \treesof{\vocab}$, and if
$t_1,\ldots,t_k \in \treesof{\vocab}$ and $f $ is an arity-$k$ symbol
in $\vocab$, for $k>0$, then $f(t_1,\ldots,t_k) \in \treesof{\vocab}$.
A {\em cost grammar} $\CG$ is defined as a tuple $(\vocab, \trees)$
where $\vocab$ is a ranked alphabet and $\trees$ is a {\em regular}
subset of $\treesof{\vocab}$. In this paper, we define this regular
subset using a grammar containing a single nonterminal.  In
particular, we focus on the following grammars: For a binary function
symbol $+$, the terms of the {\it additive-grammar $\CG(+)$} are
specified by $t := +(t,t) \sep c$, where $c$ is a constant, and the
terms of the {\it increment-grammar $\CG(+c)$} are specified by $t :=
+(t,c) \sep c$.  Given binary functions $\min$ and $+$, the terms of
the {\em min-inc-grammar $\CG(\min,+c)$\/} are given by $t :=
\min(t,t) \sep +(t,c) \sep c$, which restricts the use of addition
operation.  Given binary functions $+$ and $*$, the terms of the {\it
inc-scale Grammar $\CG(+c,*d)$} are generated by the left-linear
grammar $t := +(t,c) \sep *(t,d) \sep c$, that uses both operations in
a restricted manner, and $c$ and $d$ denote constants, ranging over
possibly different subsets of domain elements.

\mypar{Cost Models}
Given a cost grammar $\CG = (\vocab, \trees)$, a cost model
$\CostModel$ is defined as the tuple $(\CG, \domain, \interp{.})$,
where the {\em cost domain\/} $\domain$ is a finite or infinite set.
For each constant $c$ in $\vocab$, $\interp{c}$ is a unique value in
the domain $\domain$, and for each function symbol $f$ of arity $k$,
$\interp{f}$ defines a function $\interp{f}:\domain^k \mapsto\domain$.
We can inductively extend the definition of $\interp{.}$ to assign
semantics to the terms in $\trees$ in a standard fashion.  For a
numerical domain $\domain$ such as $\Nat$ (the set of natural numbers)
and $\mathbb{Z}$ (the set of integers), we use $\CostModel(\domain,+)$
to denote the cost model with the cost grammar $\CG(+)$, domain
$\domain$, and $\interp{+}$ as the standard addition operation.
Similarly, $\CostModel(\Nat,+c)$ denotes the cost model with the cost
grammar $\CG(+c)$, domain $\Nat$, and $\interp{+}$ as the standard
addition operation.  For cost grammars with two operations we denote
the cost model by listing the domain, the two functions, and sometimes
the subdomain to restrict the set of constants used by different
rules: $\CostModel(\posrat,+c,[0,1],*d)$ denotes the cost model with
the cost grammar $\CG(+c,*d)$, the set $\mathbb{Q}^{+}$ of
non-negative rational numbers as the domain, $+$ and $*$ interpreted
as standard addition and multiplication, and the rational numbers in
the interval $[0,1]$ as the range of encodings corresponding to the
scaling factor $d$.

\mypar{Cost Register Automata}
A {\em cost register automaton} (\EDWA) is a deterministic machine
that maps strings over an input alphabet to cost values using a
finite-state control and a finite set of cost registers.  At each
step, the machine reads an input symbol, updates its control state,
and updates its registers using a parallel assignment.  It is
important to note that the machine does not test the values of
registers, and thus the registers are used in a ``write-only'' mode.
The definition of such a machine is parameterized by the set of
expressions that can be used in the assignments.  Given a set $X$ of
registers and a cost grammar $\CG$, we define the set of assignment
expressions $\edwaexpr{\CG,X}$ by extending the set of terms in $\CG$
so that each internal node can be replaced by a register name.  For
example, for the additive grammar $\CG(+)$, we get the set
$\edwaexpr{+,X}$ of expressions defined by the grammar $e :=
+(e,e)\sep c\sep x$, for $x\in X$; and for the $\CG(+c,*d)$, we get
the expressions $\edwaexpr{+c,*d,X}$ defined by the grammar $e :=
+(e,c)\sep *\!(e,d)\sep c\sep x$, for $x\in X$.  We assume that the
ranked alphabet contains a special constant symbol, denoted 0, used as
the initial value of the registers.

Formally, a cost register automaton $\edwa$ over a cost grammar $\CG$ is a tuple
$(\inputalph, \edwastates, \edwainitst, \edwavariables,
\edwatrans, \edwavarup, \edwafinal)$ where
$\inputalph$ is a finite input alphabet,
$\edwastates$ is a finite set of states,
$\edwainitst\in\edwastates$ is the initial state,
$\edwavariables$ is a finite set of registers,
$\edwatrans: \edwastates\times\inputalph\mapsto\edwastates$ is the state-transition function,
$\edwavarup: \edwastates\times\inputalph\times\edwavariables\mapsto\edwaexpr{\CG,\edwavariables}$
is the register update function, and
$\edwafinal : \edwastates \mapsto \edwaexpr{\CG,\edwavariables}$ is a {\em partial\/}
final cost function.

The semantics of such an automaton is defined with respect to a cost
model $\CostModel=(\CG, \domain, \interp{.})$, and is a partial
function $\computation{\edwa,\CostModel}$ from $\fm\inputalph$ to
$\domain$.  A configuration of $\edwa$ is of the form
$(\edwastate,\edwavaluation)$, where $\edwastate\in\edwastates$ and
the function $\valuation:\edwavariables\mapsto\domain$ maps each
register to a cost in $\domain$.  Valuations naturally map expressions
to cost values using the interpretation of function symbols given by
the cost model.  The initial configuration is
$(\edwainitst,\valuation_0)$, where $\valuation_0$ maps each register
to the initial constant $0$.  Given a string $w=a_1\ldots
a_n\in\fm\inputalph$, the run of $\edwa$ on $w$ is a sequence of
configurations $(\edwastate_0,\valuation_0)\ldots
(\edwastate_n,\valuation_n)$ such that for $1\le i \le n$,
$\edwatrans(\edwastate_{i-1},a_i)=\edwastate_{i}$ and for each $x \in
\edwavariables$, $\valuation_{i}(x) =
\interp{\valuation_{i-1}(\edwavarup(\edwastate_{i-1},a,x))}$.  The
output of $\edwa$ on $w$, denoted by
$\computation{\edwa,\CostModel}(w)$, is undefined if
$\edwafinal(\edwastate_n)$ is undefined, and otherwise it equals
$\interp{\valuation_n(\edwafinal(\edwastate_n))}$.

\mypar{CRA-definable Cost Functions}
Each cost model $\CostModel=(\CG, \domain, \interp{.})$ defines a
class of cost functions $\CF(\CostModel)$: a partial function $f$ from
$\fm\inputalph$ to $\domain$ belongs to this class iff there exists a
\EDWA\ $\edwa$ over the cost grammar $\CG$ such that $f$ equals
$\computation{\edwa,\CostModel}$.  The class of cost functions
corresponding to the cost model $\CostModel(\Nat,+)$ is abbreviated as
$\CF(\Nat,+)$, the class corresponding to the cost model
$\CostModel(\posrat,+c,[0,1],*d)$ as $\CF(\posrat,+c,[0,1],*d)$, etc.

\newcommand{\sqp}{\ensuremath{\!\!+\!\!}}
\newcommand{\muc}{\multicolumn{1}{c}}

\begin{figure}[t]
\begin{tikzpicture}[->,-stealth',semithick]

\node[state] (q0) {$q_0$};
\node[coordinate,node distance=7mm,above left of=q0] (start1) {};
\draw (start1) to (q0);

\draw (q0) to[loop left] node[smalltext, left] (t1) {} (q0);
\node[smalltext,node distance=5mm,left of=t1]
     {$\begin{array}{c}
              a/ \\
              x := x \sqp 1 \\
              y := y \sqp 1
     \end{array}$};

\draw (q0) to[loop above] node[smalltext, above] (t2)
        {$b\hspace{-.5em}\left/\hspace{-.8em}
         \begin{array}{l} x := x \\
                          y := y \sqp 1
         \end{array}\right.\hspace{-.5em}$}
      (q0);

\draw (q0) to[loop below] node[smalltext, below] (t3)
        {$e\hspace{-.5em}\left/\hspace{-.8em}
          \begin{array}{l} x := y\sqp 1 \\
                           y := y \sqp 1
          \end{array}\right.\hspace{-.5em}$}
      (q0);

\node[smalltext,node distance=19mm,below of=q0] {$\mu(q_0) = x$};

\node[coordinate,node distance=27mm,below of=q0]  (co1) {};
\node[captiontext,node distance=3mm, left of=co1] (cap1) {$M_1$: CRA over};
\node[coordinate,node distance=3mm,below of=cap1]  (co11) {};
\node[captiontext, node distance=3.5mm,right of=co11] {$(+c)$};

\node[state,right of=q0,node distance=30mm] (q02) {$q_0$};
\node[coordinate,node distance=7mm,above left of=q02] (start2) {};
\node[font=\fontsize{6}{6}\selectfont,right of=start2,node distance=1.5mm,
      anchor=south east] {$x\!\!:=\!\!\infty$};
\draw (start2) to (q02);

\draw (q02) to[loop left] node[smalltext, left, node distance=1.5mm] (t4) {}
      (q02);
\node[smalltext,node distance=6mm,left of=t4]
     {$\begin{array}{l} \muc{a/} \\
                        x := x \\
                        y := y \sqp 1\\
                        z := z \\
       \end{array}$};
\draw (q02) to[loop above] node[smalltext, above] (t5) {}
      (q02);

\node[smalltext,node distance=6mm,above right of=t5]
      {$b\hspace{-.5em}\left/\hspace{-.5em}
        \begin{array}{l} x := x \\
                         y := y \\
                         z := z \sqp 1
        \end{array}\right.$};

\draw (q02) to[loop below] node[smalltext, below, node distance=1.5mm]
        {$e\hspace{-.5em}\left/\hspace{-.5em}
          \begin{array}{l} x := min(x,y,z) \\
                           y := 0 \\
                           z := 0
          \end{array}\right.$}
      (q02);

\node[smalltext,node distance=19mm,below of=q02] {$\mu(q_0) = x$};

\node[coordinate,node distance=27mm,below of=q02]  (co2) {};
\node[captiontext,node distance=3mm, left of=co2] (cap1) {$M_2$: Copyless CRA};
\node[coordinate,node distance=3mm,below of=cap1]  (co22) {};
\node[captiontext, node distance=3.5mm,right of=co22] {over $(min,+c)$};

\node[state,right of=q02,node distance=22mm] (q03) {$q_0$};
\node[state,right of=q03,node distance=25mm] (q13) {$q_1$};
\node[coordinate,node distance=7mm,above left of=q03] (start3) {};
\draw (start3) to (q03);
\draw (q03) to[loop below] node[smalltext, below, node distance=1.5mm]
        {$\begin{array}{l} \muc{a/} \\
                          x := x \sqp 1 \\
                          y := y
         \end{array}$}
      (q03);
\draw (q03) to[loop above] node[smalltext, above]
        {$\begin{array}{l} \muc{b/} \\
                           x := x \\
                           y := y
          \end{array}$}
      (q03);
\draw (q03) to node[node distance=2mm,above,smalltext] (mid)
        {$e\hspace{-.5em}\left/\hspace{-.5em}
          \begin{array}{l} x := x \\
                           y := x
          \end{array}\right.$}
      (q13);
\draw (q13) to[loop above] node[smalltext, above]
        {$\begin{array}{l}   \muc{a/} \\
                             x := x \sqp 1 \\
                             y := y
          \end{array}$}
      (q13);

\draw (q13) to[loop right] node [smalltext, right] (lab3) {} (q13);
\node[smalltext,right of=lab3,node distance=4mm]
     {$\begin{array}{l} \muc{b/} \\
                        x := x \\
                        y := y \sqp 1
       \end{array}$};
\draw (q13) to[loop below] node[smalltext, below, node distance=1.5mm]
        {$\begin{array}{l} e/ \\
                           x := x \\
                           y := min(x,y)
          \end{array}$}
      (q13);
\node[smalltext,node distance=24mm,below of=mid]
    {$\begin{array}{l}\mu(q_0) = x \\
                      \mu(q_1) = y
      \end{array}$};
\node[captiontext,node distance=31mm, below of=mid] (cap3) {$M_3$: CRA over};
\node[coordinate,node distance=3mm,below of=cap3] (co3) {};
\node[captiontext,node distance=4mm, right of=co3] {$(min,+c)$};

\node[state,right of=q03,node distance=52mm] (q04) {$q_0$};
\node[state,right of=q04,node distance=20mm] (q14) {$q_1$};
\node[coordinate,node distance=7mm,above left of=q04] (start4) {};
\draw (start4) to (q04);
\draw (q04) to[loop above] node [smalltext,above] {$a/x := x \sqp 10$} (q14);
\draw (q04) to[loop below] node [smalltext,below] {$e/x := x$} (q14);
\draw (q04) to node[smalltext,above] (mid4) {$b/x:=x$} (q14);
\draw (q14) to[loop below] node[smalltext,below]
     {$\begin{array}{c} e/ \\
                        x:= 0.95*x
       \end{array}$}
      (q14);
\draw (q14) to[loop above] node[smalltext,above]
     {$\begin{array}{c} a,b/ \\
                        x:=x
       \end{array}$}
      (q14);
\node[smalltext,node distance=23mm,below of=mid4]
    {$\begin{array}{l}\mu(q_0) = x \\
                      \mu(q_1) = x
      \end{array}$};
\node[captiontext,node distance=30mm,below of=mid4] (cap4) {$M_4$: CRA over};
\node[coordinate,node distance=3mm,below of=cap4] (co4) {};
\node[captiontext,node distance=3mm, right of=co4] {$(+c,*d)$};

\end{tikzpicture}
\caption{Examples of Cost Register Automata\label{rcfex1}}
\end{figure}
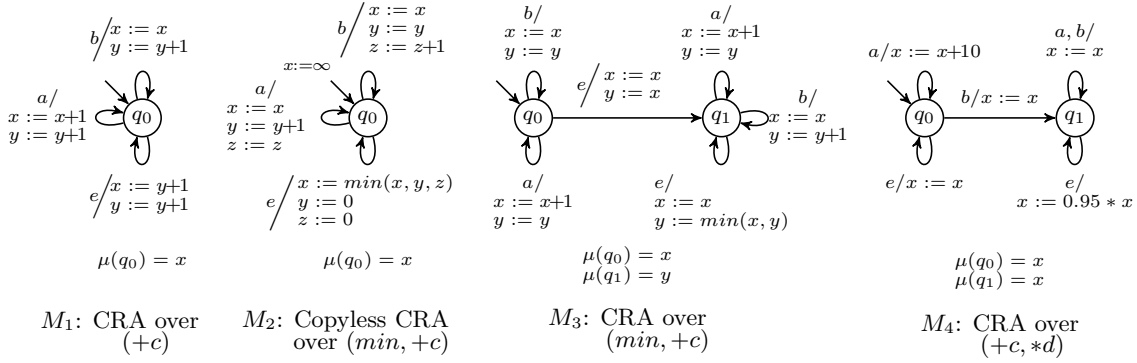

\mypar{Examples}
Figure~\ref{rcfex1} shows examples of cost register automata for
$\Sigma=\{a,b,e\}$.  Consider the cost function $f_1$ that maps a
string $w$ to the length of the substring obtained by deleting all
$b$'s after the last occurrence of $e$ in $w$.  The automaton
$\edwa_1$ computes this function using two cost registers and
increment operation.  The register $y$ is incremented on each symbol,
and hence equals the length of string processed so far.  The register
$x$ is not incremented on $b$ symbols, but is updated to the total
length stored in $y$ when $e$ symbol is encountered.  This example
illustrates the use of two registers: the computation of the desired
function $f_1$ in register $x$ update crucially relies on the
auxiliary register $y$.

For a string $w$ and symbol $a$, let $|w|_a$ denote the count of $a$
symbols in $w$.  For a given string $w$ of the form $w_1\,e\,w_2\ldots
e\,w_{n-1}\,e\,w_n$, where each block $w_i$ contains only $a$'s and
$b$'s, let $f_2(w)$ be the minimum of the set
$\{|w_{n-1}|_a,|w_{n-1}|_b,|w_n|_a,|w_n|_b\}$.  The \EDWA\ $\edwa_2$
over the grammar $\CG(\min,+c)$ computes this function using three
registers by an explicit application of the $\min$ operator.

For a given string $w=w_1\,e\, w_2\, e \ldots e\, w_n$, where each
$w_i$ contains only $a$'s and $b$'s, consider the function $f_3$ that
maps $w$ to $\min_{j=1}^{n-1} (|w_1|_a + |w_2|_a + \cdots |w_j|_a +
|w_{j+1}|_b + \cdots + |w_n|_b)$.  This function is computed by the
\EDWA\ $\edwa_3$ over the grammar $\CG(\min,+c)$.

The final example concerns use of scaling. Consider a computation
where we wish to charge a cost of $10$ upon seeing an $a$ event until
a $b$ event occurs.  Once a $b$ event is triggered, for every
subsequent $e$ event, the cost is discounted by $5\%$. Such a cost
function is computed by the \EDWA\ $\edwa_4$ over the grammar
$\CG(+c,*d)$.

\mypar{Copyless Restriction}
A \EDWA\ $\edwa$ is said to be {\em copyless\/} if each register is
used at most once at every step: for each state $\edwastate$ and input
symbol $a$ and each register $x\in X$, the register $x$ appears at
most once in the set of expressions $\{\edwatrans(\edwastate,a,y)\sep
y \in X\}$ and $x$ appears at most once in the output expression
$\edwafinal(\edwastate)$.  Each cost model $\CostModel=(\CG, \domain,
\interp{.})$ then defines another class of cost functions
$\CCF(\CostModel)$: a partial function $f$ from $\fm\inputalph$ to
$\domain$ belongs to this class iff there exists a copyless \EDWA\
$\edwa$ over the cost grammar $\CG$ such that $f$ equals
$\computation{\edwa,\CostModel}$.  In Figure~\ref{rcfex1}, the
automata for function $f_2$ and $f_4$ are copyless, while the ones for
$f_1$ and $f_3$ are not.

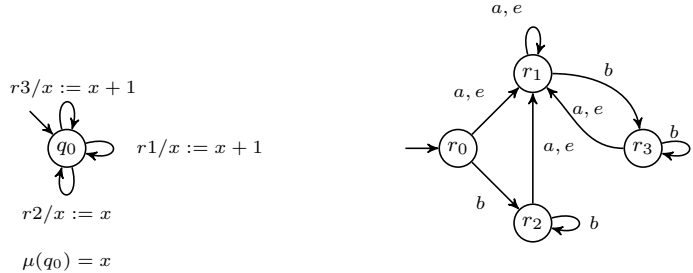
\begin{figure}[t]
\centering
\begin{tikzpicture}[->,-stealth',semithick]

\node[state] (q0) {$q_0$};
\node[coordinate,node distance=7mm,above left of=q0] (start0) {};

\draw (start0) to (q0);

\draw (q0) to[loop above] node[smalltext,above,near end]
           {$\begin{array}{r} r3/
                              x:=x+1
             \end{array}$} (q0);
\draw (q0) to[loop below] node[smalltext,below]
           {$\begin{array}{l} r2/
                              x:=x
             \end{array}$}  (q0);
\draw (q0) to[loop right] node[smalltext,right]
           {$\begin{array}{l} r1/
                              x:=x+1
             \end{array}$} (q0);
\node[smalltext,below of=q0,node distance=15mm] {$\mu(q_0) = x$};

\node[state,right of=q0,node distance=52mm] (r0) {$r_0$};
\node[state,above right of=r0,node distance=40] (r1) {$r_1$};
\node[state,below right of=r0,node distance =40] (r2) {$r_2$};
\node[state,right of=r0,node distance=70] (r3) {$r_3$};
\node[coordinate,node distance=7mm,left of=r0] (start1) {};

\draw (start1) to (r0);
\draw (r0) to node [smalltext,above left] {$a,e$} (r1);
\draw (r0) to node [smalltext,below left] {$b$} (r2);
\draw (r1)[loop above] to node [smalltext,above left] {$a,e$} (r1);
\draw (r1) to[out=0, in=100] node [smalltext,above] {$b$} (r3);
\draw (r2)[loop right] to node [smalltext,right] {$b$} (r2);
\draw (r2) to node [smalltext,right] {$a,e$} (r1);
\draw (r3)[loop right] to node [smalltext,above left] {$b$} (r3);
\draw (r3) to[out=180, in=315] node [smalltext,above] {$\ \ a,e$} (r1);


\end{tikzpicture}
\caption{On the left a \REDWA over $(+c)$ corresponding to $M_1$ in
Figure~\ref{rcfex1}. On the right the corresponding labeling automaton.
The states of $A$ corresponds to the languages used in the informal description of $M_1$.
\label{cra_rla}}
\end{figure}

\mypar{Regular Look-Ahead}\label{subsec:rla}
A \EDWA\ $\edwa^R$ with {\em regular look-ahead} (\REDWA) is a
\EDWA\xspace that can make its decisions based on whether the
remaining suffix of the input word belongs to a regular language.  Let
$L$ be a regular language, and let $A$ be a DFA for $reverse(L)$ (such
a DFA exists, since regular languages are closed under the reverse
operation). Then, while processing an input word, testing whether the
suffix $a_j\ldots a_k$ belongs to $L$ corresponds to testing whether
the state of $A$ after processing $a_k\ldots a_j$ is an accepting
state of $A$.  We now try to formalize this concept Let $w = a_1\ldots
a_k$ be a word over $\Sigma$, and let $A$ be a DFA with states $R$
processing words over $\Sigma$. Then the \emph{$A$-look-ahead labeling} of
$w$, is the word $w_A = r_1r_2\ldots r_k$ over the alphabet $R$ such
that for each position $1 \leq j \leq k$, the corresponding symbol is
the state of the DFA $A$ after reading $a_k\ldots a_j$ (it reads the
reverse of the word).  A \REDWA consists of an DFA $A$ over
$\Sigma$ with states $R$, and a \EDWA $\edwa$ over the input
alphabet $R$.  The output of \REDWA $(\edwa,A)$ on $w$, denoted by
$\computation{(\edwa,A),\CostModel}(w)$, is defined as
$\computation{\edwa,\CostModel}(w_A)$.  In Figure~\ref{cra_rla} we
show the \REDWA for $M_1$ of Figure~\ref{rcfex1}.

\section{Regular Cost Functions} \label{sec:rcf}
Consider a cost grammar $\CG=(\vocab,\trees)$.  The terms in $\trees$
can be viewed as trees: an internal node is labeled with a function
symbol $f$ of arity $k>0$ and has $k$ children, and each leaf is
labeled with a constant.  A {\em deterministic streaming
string-to-tree transduction\/} is a (partial) function $f:\fm{\Sigma}
\mapsto \trees$.  The theory of such transductions has been well
studied, and in particular, the class of {\em regular\/}
string-to-tree transductions has appealing closure properties, and
multiple characterizations using Macro-tree-transducers (with
single-use restriction and regular
look-ahead)~\cite{engelfriet_macro2_1999}, Monadic-Second-Order logic
definable graph transformations~\cite{courcelle_graph_2002}, and
streaming tree transducers~\cite{alur_stt_2011}.  We first briefly
recap the model of streaming string-to-tree transducers.

\mypar{Streaming String-to-Tree Transducers (\SSTT)} A streaming
string-to-tree transducer is a deterministic machine model that can
compute regular transformations from strings to ranked trees in a
single pass.  We note that \SSTT can be viewed as a variant of
\EDWA~\footnote{To make the connection precise, we need to allow
registers in CRAs to be {\em typed\/}, and use function symbols with
typed signatures. We also use $\mu$ to denote the output function
instead of $F$,. For simplicity of presentation, we defer this detail
to a later version.}, where each register stores a term, that is, an
{\em uninterpreted expression}, and these terms are combined using the
rules allowed by the grammar.  To obtain a model whose expressiveness
coincides with the regular transductions, we must require that the
updates are copyless, but need to allow terms that contain ``holes'',
\ie, parameters that can be substituted by other terms.

Let $\CG=(\vocab,\trees)$ be a cost grammar.  Let $\hole$ be a special
0-ary symbol that denotes a place holder for the term to be
substituted later. We obtain the set $\ptrees$ by adding the symbol
$\hole$ to $\vocab$, and requiring that each term has at most one leaf
labeled with $\hole$.  For example, for the cost grammar $\CG(+c)$,
the set $\ptrees$ of parameterized terms is defined by the grammar $t
:= +(t,c)\sep c\sep \hole$; and for the cost grammar $\CG(+)$, the set
$\ptrees$ of parameterized terms is defined by the grammar $t' :=
+(t,t')\sep +(t',t) \sep c\sep \hole$, where $t$ stands for (complete)
terms generated by the original grammar $t := +(t,t)\sep c$.  A
parameterized term such as $\min(5,\hole+3)$ stands for an incomplete
expression, where the parameter $\hole$ can be replaced by another
term to complete the expression.  Registers of an \SSTT hold
parameterized terms.  The expressions used to update the registers at
every step are given by the cost grammar, with an additional rule for
{\em substitution\/}: given a parameterized expression $e$ and another
expression $e'$, the expression $e[e']$ is obtained by substituting
the sole $\hole$-labeled leaf in $e$ with the expression $e'$.

Given a set $\edwavariables$ of registers, the set
$\pexpr(\CG,\edwavariables)$ represents parameterized expressions that
can be obtained using the rules of $\CG$, registers in
$\edwavariables$, and substitution.  For example, for the grammar
$\CG(+c)$ and a set $\edwavariables$ of registers, the set
$\pexpr(+c,\edwavariables)$ is defined by the grammar $e := +(e,c)\sep
c\sep \hole \sep x \sep e[e]$, for $x\in\edwavariables$.  The output
of an \SSTT  is a (complete) term in $\trees$ defined using the final
cost function.  The register update function and the final cost
function are required to be {\em copyless}: each register is used at
most once on the right-hand-side in any transition.  The semantics of
an \SSTT gives a partial function from $\fm\inputalph$ to $\trees$.
We refer the reader to \cite{alur_stt_2011} for details.

\mypar{Regular Cost Functions}\label{subsec:reg}
Let $\Sigma$ be a finite input alphabet.  Let $\domain$ be a cost
domain. A {\em cost function} $f$ maps strings in $\fm{\Sigma}$ to
elements of $\domain$. Let $\CostModel = (\CG, \domain, \interp{.})$
be a cost model.  A cost function $f$ is said to be {\em regular with
respect to the cost model $\CostModel$\/} if there exists a regular
string-to-tree transduction $g$ from $\fm\Sigma$ to $\trees$ such that
for all $w \in \fm{\Sigma}$, $f(w) = \interp{g(w)}$.  That is, given a
cost model, we can define a cost function using an \SSTT: the \SSTT
maps the input string to a term, and then we evaluate the term
according to the interpretation given by the cost model.  The cost
functions obtained in this manner are the regular functions.  We use
$\reg{\CostModel}$ to denote the class of cost functions regular with
respect to the cost model $\CostModel$.

As an example, suppose $\Sigma=\{a,b\}$.  Consider a vocabulary with
constant symbols $0$, $c_a$ and $c_b$, and the grammar $\CG(+c)$.
Consider the \SSTT $\stt$ with a single register that is initialized
to $0$, and at every step, it updates $x$ to $+(x,c_a)$ on input $a$,
and $+(x,c_b)$ on input $b$.  Given input $w_1\ldots w_n$, the \SSTT
generates the term $e=+(\cdots +(+(0,c_1),c_2)\cdots c_n)$, where each
$c_i=c_a$ if $w_i=a$ and $c_i=c_b$ otherwise.  To obtain the
corresponding cost function, we need a cost model that interprets the
constants and the function symbol $+$, and we get the cost of the
input string by evaluating the expression $e$.  Now, consider another
\SSTT $\stt'$ that uses a single register initialized to $\hole$.  At
every step, it updates $x$ to $x[+(\hole,c_a)]$ on input $a$, and
$x[+(\hole,c_b)]$ on input $b$, using the substitution operation.  The
output is the term $x[0]$ obtained by replacing the parameter by 0.
Given input $w_1\ldots w_n$, the \SSTT generates the term $e'=+(\cdots
+(0,c_n),\cdots c_1)$, where each $c_i=c_a$ if $w_i=a$ and $c_i=c_b$
otherwise.  Note that the \SSTT~$\stt$ builds the cost term by adding
costs on the right, while the \SSTT~$\stt'$ uses parameter
substitution to build costs terms in the reverse order.  If the
interpretation of the function $+$ is not commutative, then these two
mechanisms allow to compute different functions, both of which are
regular.

Let's now consider a different grammar. Consider the constant symbols
$1$, $c_a$ and $c_b$, and the grammar $\CG(+c,*d)$.
Consider the \SSTT~$\stt$ with a single register that is initialized
to $1$ and, at every step, it updates $x$ to $+(*(x,c_a),1)$ on input $a$
and to $+(*(x,c_b),1)$ on input $b$.
Given input $w_1\ldots w_n$, the \SSTT
generates the term
$e=+(*(+(*(\cdots +(*(1,c_b),1) \cdots,c_{n-1}),1),c_n),1)$, where each
$c_i=c_a$ if $w_i=a$ and $c_i=c_b$ otherwise.
Now, consider another
\SSTT $\stt'$ that uses a single register initialized to $\hole$.
At every step, it updates $x$ to $x[+(*(\hole,c_a),1)]$ on input $a$, and
$x[+(\hole,c_b)]$ on input $b$, using the substitution operation.  The
output is the term $x[1]$ obtained by replacing the parameter by 1.
Given input $w_1\ldots w_n$, the \SSTT
generates the term
$e=+(*(+(*(\cdots +(*(1,c_b),1) \cdots,c_{2}),1),c_1),1)$, where each
$c_i=c_a$ if $w_i=a$ and $c_i=c_b$ otherwise. The same considerations as before follow.

\mypar{Closure Properties}
If $f$ is a regular cost function from $\fm{\Sigma}$ to a cost domain
$\domain$, then the {\em domain} of $f$, \ie, the set of strings $w$
such that $f(w)$ is defined, is a regular language. Closure properties
for regular string-to-tree transductions immediately imply certain
closure properties for regular cost functions.  For a string $w$, let
$w^{r}$ denote the reverse string.  We define a {\em reverse function}
$f^{r}:\fm{\Sigma} \mapsto \domain$ such that for all $w \in
\fm{\Sigma}$, $f^{r}(w) = f(w^r)$.
Given cost functions $f_1,f_2$ from $\fm{\Sigma}$ to $\domain$ and a
language $L\subseteq \fm\Sigma$, the {\em choice\/} function ``if $L$
then $f_1$ else $f_2$'' maps an input string $w$ to $f_1(w)$ if $w\in
L$, and to $f_2(w)$ otherwise.  If the two cost functions $f_1$ and
$f_2$ are regular and if $L$ is a regular language, then the choice
function is also regular. We now show that regular cost functions are
closed under reverse and regular choice.

\begin{theorem}[Closure Properties of Regular Cost Functions]
For every cost model $\CostModel$,
\begin{enumerate}
\item[(a)]
if a cost function $f$ belongs to the class $\reg{\CostModel}$, then
so does the function $f^r$;
\item[(b)]
if cost functions $f_1$ and $f_2$ belong to the class $\reg{\CostModel}$,
then so does the function ``if $L$ then $f_1$ else $f_2$'' for every regular language $L$.
\end{enumerate}
\end{theorem}

\Proof~ (a) The proof of the first statement follows from the
application of theorems from
\cite{alur_expressiveness_2010,alur_stt_2011}.  Let $f$ be a regular
cost function belonging to the class $\reg{\CostModel}$, and let
$\stt$ be an \SSTT that computes the function $f$. A streaming string
transducer (\SST) is a machine similar to an \SSTT. It maps strings to
strings with the help of a fixed number of registers that store
strings, and uses updates that involve concatenating registers and
strings in a copyless fashion.  Given an alphabet $\Sigma$, computing
the reverse $w^r$ of strings $w\in \Sigma$ is an \SST-definable
transduction \cite{alur_expressiveness_2010}.  Let \SST~$\stt'$ that
computes such a transduction. As proved in \cite{alur_stt_2011}, the
composition of an \SST-definable transduction and an \SSTT~definable
transduction is an \SSTT~definable transduction.  Thus the sequential
composition of $\stt'$ and $\stt$ is also \SSTT~definable. Thus, for
every regular function $f$ in $\reg{\CostModel}$, we can construct the
\SSTT $\stt \circ \stt'$ that maps every string $w$ to the
corresponding term $f(w^r)$, which is, by definition, the cost
function $f^r(w)$.   Thus, if $f$ belongs to $\reg{\CostModel}$, so
does $f^r$.

(b) To prove the second statement, we show how we can construct an
\SSTT $\stt_c$ that defines the function ``if $L$ then $f_1$ else
$f_2$.''  As $f_1$ and $f_2$ are regular cost functions, they are
definable by \SSTT~$\stt_1$ and $\stt_2$ respectively. The
\SSTT~$\stt_c$ maintains $2$ disjoint sets of registers, where the
first set corresponds to registers of $\stt_1$ and the second to the
registers of $\stt_2$. A state of $\stt_c$ is a tuple $(q,q_1,q_2)$,
where $q_1$ and $q_2$ exactly track the states of $\stt_1$ and
$\stt_2$, and $q$ is the state of the DFA corresponding to the regular
language $L$. The output function of $\stt_c$ is defined such that if
the input word $w$ is in $L$ (\ie, it is accepted by the corresponding
DFA), then $\stt_c$ uses the first set of registers to compute the
output, and, otherwise uses the second set of registers.  For any cost
model, the $\stt_c$ exactly defines the choice function.  \qed

An \SSTT with regular look ahead is a pair $(\stt,A)$ where $\stt$ is
an \SSTT and $A$ a DFA.  As discussed earlier regular-look-ahead tests
allow machines to make its decisions based on whether the remaining
suffix of the input word belongs to a given regular language.  \SSTT
are closed under the operation of regular-look-ahead
\cite{alur_stt_2011}, which implies the same for regular cost
functions.

\begin{theorem}[Closure Under RLA]\label{thm:stt-rla}
For every cost model \SSTT with regular-look-ahead $(\stt,A)$, there
exists an \SSTT $\stt'$ without regular-look-ahead which computes the
same function.
\end{theorem}
\Proof~Theorem 9 in \cite{alur_stt_2011}.
\qed

\mypar{Constant Width and Linear Size of Output Terms}
While processing symbols of an input string $w$, in each step an \SSTT
performs a copyless update. Thus, the sum of the sizes of all terms
stored in registers grows only by a constant additive factor.  It
follows that $|\stt(w)|$ is $O(|w|)$.  Viewed as a tree, the depth of
$\stt(w)$ can be linear in the length of $w$, but its width is
constant, bounded by the number of registers.  This implies that if
$f$ is a cost function in $\reg{\Nat,+c}$, then $|f(w)|$ must be
$O(|w|)$. In particular, the function $f(w)=|w|^2$ is not regular in
this cost model.  Revisiting the examples in \secref{cra}, it turns
out that the function $f_1$ is regular for $\CostModel(\Nat,+c)$, and
the function $f_2$ is regular for $\CostModel(\Nat,\min,+c)$.  The
function $f_3$ does not appear to be
regular for $\CostModel(\Nat,\min,+c)$, as it seems to require
$O(|w|^2)$ terms to construct it.

\section{Commutative-Monoid Cost Functions}\label{sec:cmcf}
In this section we explore and analyze cost functions for the cost
models of the form $(\domain,\myplus)$, where $\domain$ is a cost
domain (with a designated identity element) and the interpretation
$\interp{\myplus}$ is a commutative and associative function.

\mypar{Expressiveness}
Given a cost model $(\domain,\myplus)$, we can use regular
string-to-term transductions to define two (machine-independent)
classes of functions: the class $\reg{\domain,\myplus c}$ defined by
the grammar $\CG(\myplus c)$ and the class $\reg{\domain,\myplus}$
defined by the grammar $\CG(\myplus)$.  Relying on commutativity and
associativity, we show these two classes to be equally expressive.
This class of ``regular additive cost functions'' corresponds exactly
to functions computed by CRAs with increment operation, and also, by
copyless-CRAs with addition.  We are going to show the result via
intermediate results.

\begin{lemma}
\label{ssttostt}
For cost domain $\domain$ with a commutative and associative operation
$\interp{\myplus}$,
$\reg{\domain,\myplus c}=\reg{\domain,\myplus}$.
\end{lemma}
\Proof~ $\reg{\domain,\myplus c} \subseteq \reg{\domain,\myplus}$ is
trivially true.  We now prove the other direction.  Consider a
function $f:\Sigma\st\mapsto\domain$ over $\reg{\domain,\myplus}$
defined by the \SSTT~$\stt$. In effect, $\stt$ outputs trees over the
grammar $G(\myplus)$.  We show that there exists an \SSTT~$\stt'$ that
outputs trees over the grammar $G(\myplus c)$ and computes $f$.  As
shown in \cite{alur_stt_2011} the class of string-to-tree
transductions computable by an \SSTT coincides with the class of
transductions computed by MSO transducers
\cite{engelfriet_macro2_1999}. Thus, for a given \SSTT $\stt$, there
is an MSO string-to-tree transducer $M$ that computes $f$.

The {\em yield} of a tree $t$ is the string obtained by concatenating all
the leaves of $t$ as they appear in an pre-order search.  In Lemma 7.6
of \cite{engelfriet_macro2_1999}, the authors prove that computing the
yield of a tree is an MSO-definable transduction.  Consider the MSO
tree-to-string transducer $M'$ that computes the yield of trees over
the grammar $G(\myplus)$.  Essentially, the yield of a tree $t$
generated by the \SSTT $\stt$ contains constant symbols. If
$\interp{\myplus}$
is a commutative and associative operator, the order in which the
symbols appear is not important, and any tree $t'$ with internal nodes
$\myplus$ and leaves corresponding to the yield of $t$ represents an
equivalent expression to the one represented by $t$.

$M'$ outputs strings over $C$ where $C$ is the set of constants in the
output alphabet of $M$.  Given a string $c_1\ldots c_n$ over $C$ we
want to produce the string $c_1\myplus
(c_2\myplus(\ldots(c_n\myplus(0))))$ that belongs to the grammar
$G(\myplus c)$. This transduction is clearly MSO-definable (it is a
simple relabeling). Let $M''$ be the MSO transducer that computes this
relabeling.

As MSO transducers are closed under composition
\cite{engelfriet_macro2_1999} the transduction $M_f=M\circ M'
\circ M''$ is also MSO-definable.  Observe that for any string $w
\in \fm\Sigma$, $M_f(w)$ is a term equivalent to the function $f(w)$,
but is a term over the cost grammar $\CG(\myplus c)$ (due to
associativity and commutativity of $\interp{\myplus}$).  Clearly,
$M_f$ computes the function $f$.  As MSO-definable transductions are
equivalent to \SSTT-definable transductions, there is an \SSTT
$\stt'$ equivalent to $M_f$. Thus for every $f$ in
$\reg{\domain,\myplus}$ (defined by the \SSTT $\stt$) there is an
\SSTT $\stt'$ that defines an equivalent function in
$\reg{\domain,\myplus c}$.
\qed

\begin{lemma}
\label{mvcmrtosst}
For cost domain $\domain$ with a commutative associative operation $\myplus$,
$\CF(\domain,\myplus c)\subseteq \reg{\domain,\myplus c}$.
\end{lemma}
\Proof~ Consider a cost function $f:\Sigma\st\mapsto\domain$ belonging
to $\CF(\domain,\myplus c)$.  We show that we can construct an \SSTT
$U$ such that for all $w \in \fm\Sigma$, $\interp{U}(w) = f(w)$.

Recall that a cost function in $\CF(\domain,\myplus c)$ is definable
by a \EDWA $\edwa$ over $(\domain,\myplus c)$, where $\edwa$ is given
by the tuple
$\edwa=(\inputalph,\edwastates,\edwainitst,\edwavariables,\edwatrans,\edwavarup,\edwafinal)$.
In order to construct the desired \SSTT $\stt$, we first construct an
\SSTT with regular-look-ahead, denoted by $(\stt',A)$ that computes
$f$.  Here $\stt'$ is an \SSTT defined by the tuple
$(Q,q_0,\{v\},\delta',\rho',\mu')$ and $A$ is a DFA $(R,r_0,\delta_R)$
specifying the regular-look-ahead. Given an input string $w$, recall
that $\stt'$ reads $R$-labeled words corresponding to the run of $A$
on the reverse string $w^r$.

The final cost function $\edwafinal$ of the \EDWA $\edwa$ maps a state
to a term in $E(\CG,X)$. This can be extended to the partial function
$\edwafinal\st: Q \times \Sigma\st \mapsto E(\CG,X)$ as follows. For
all $q$, $\edwafinal\st(q,\varepsilon) = \edwafinal(q)$, and
$\edwafinal\st(q,aw)$ is obtained by replacing each $x$ in
$\edwafinal\st(\delta(q,a),w)$ by the expression $\edwavarup(q,a,x)$.
$\edwafinal\st(q,w)$ gives the output of $\edwa$ starting in state $q$
after reading $w$. For the grammar $\CG(\myplus c)$, it is easy to
show by induction that for all $q$ and $w$, the expression
$\edwafinal\st(q,w)$ contains at most one register name.

We now describe how the RLA automaton $A=(R,r_0,\delta_R)$ is
constructed.  Consider an input string $w=w_1\ldots w_n$, and recall
that $A$ reads the reverse string $w^r$.  At position $i$, we need $A$
to report the register name that will contribute to the final output,
\ie, the register that ``flows'' into the final output after $\edwa$
reads the remaining suffix $w_{j+1}\ldots w_n$. To do so, each state
of $A$ is a pair  of the form $(a,\chi)$ such that
$a\in(\Sigma\cup\setof{\varepsilon})$ and
$\chi:\edwastates\mapsto\edwavariables\cup\{\varepsilon\}$ is a
function mapping every state in $\edwastates$ to a register name or a
special empty symbol.  While reading a string $w_1\ldots w_n$ in the
reverse order, the invariant maintained by a state $(w_i,\chi)$ of $A$
is that if for each state $q \in \edwastates$, if the \EDWA $\edwa$
reads the symbol $w_i$, then the register name that flows into the
final output after reading the string $w_{i+1}\ldots w_n$ is
$\chi(q)$.


The initial state of $A$, $r_0 = (\varepsilon,\chi_0)$, where
$\chi_0(q)=x$ if $x$ is the (only) register name appearing in
$\edwafinal(q)$, and is $\varepsilon$ if no register name appears in
$\edwafinal(q)$. Note that the first component of the state, \ie, the
input symbol is not used at this point as this corresponds to the case
where the \SSTT~ has reached the end of the string. We define
$\delta_R$ using the register update functions of $\edwa$ as follows:

Suppose $A$ is in state $(w_{i+1},\chi)$ and it reads the symbol
$w_i$.  We define $\delta_R((w_{i+1},\chi),w_i) = (w_i,\chi')$, where
the function $\chi'$ is defined as follows:

\[
    \forall q \in Q\ \text{s.t.}\ \delta(q,w_i) = q',\ \chi'(q) = \left\{
            \begin{array}{ll}
                \varepsilon & \text{if $\chi(q') = \varepsilon$ or if $\edwavarup(q',w_i,x) = c$}  \\
                y           & \text{if $\edwavarup(q',w_i,x) = y + c$}.
            \end{array}
                \right.
\]

In the above definition, $c$ is some constant in $\domain$.
We can now define how the state transition function $\delta'$ and
register update function $\rho'$ of $\stt'$ are defined:
$\delta'(q,(a,\chi)) = q'$ if in $\edwa$, $\delta(q,a) = q'$.  In
state $q$, $\stt'$ exactly knows the register $\chi(q)$ that
contributes to the final output by reading the symbol $(a,\chi)$.
Thus, it is enough for $\stt'$ to have just one register (denoted
$v$).  For an expression $t$ in $E(\CG,X)$, let $t[x\mapsto v]$ be the
expression obtained by renaming the register $x$ to $v$.  The register
update function $\rho'(q,(a,\chi),v)$ is defined to be
$\edwavarup(q,a,\chi(q))[v\mapsto x]$ if $\chi(q) \neq \varepsilon$,
and $0$ otherwise.

Finally, as \SSTT are closed under regular-look-ahead, there exists an
\SSTT $\stt$ equivalent to the \SSTT with regular-look-ahead $\stt'$.
Thus for every function definable by \EDWA over $(\domain, \myplus
c)$, there exists an \SSTT $\stt$ that computes $f$, which means that
$f$ is in $\reg{\domain,\myplus c}$.
\qed

\begin{lemma}
\label{ppcedwatopedwa}
For cost domain $\domain$ with a commutative associative operation
$\myplus$, $\CCF(\domain,\myplus)\subseteq \CF(\domain,\myplus c)$.
\end{lemma}

\Proof~ Consider a function $f:\Sigma\st\mapsto\domain$ belonging to
$\CCF(\domain,\myplus)$.  Let
$\edwa=(\inputalph,\edwastates,\edwainitst,\edwavariables,\edwatrans,\edwavarup,\edwafinal)$
be a copyless \EDWA~over $(\domain,\myplus)$ that computes $f$.  We
construct an
\EDWA~$\edwa'=(\inputalph,\edwastates,\edwainitst,2^\edwavariables,\edwatrans,\edwavarup',\edwafinal')$
over $(\domain,\myplus c)$ that also computes $f$.

For every subset $S\subseteq \edwavariables$, $\edwa'$ maintains a
register denoted by $x_S$.  $\edwa'$ maintains the following invariant: If
the configuration of $\edwa$ is $(\edwastate,\valuation)$, then the
corresponding configuration of $\edwa'$ is $(\edwastate,\valuation')$ such
that for all $x_S \in 2^\edwavariables$, $\valuation'(x_S)=\sum_{x\in S}
\valuation(x)$.  Informally, each register $x_S$ maintains the sum of the
registers in the set $S$.

The update function $\rho'$ of $\edwa'$ corresponding to the updates of
$\edwa$ can be defined as follows.  Let $u_S = \sum_{x\in S} \rho(q,a,x)$.
Note that the expression $u_S$ is composed of two parts: an expression
denoting the sum of register names, and a constant obtained by summing all
the constants in each of the $\rho(q,a,x)$ expressions. Let $reg(u_S)$
denote the set of registers appearing in $u_S$ and let $co(u_S)$ denote the
computed constant.  Then, we define $\rho'(q,a,x_S) = x_{reg(u_S)} +
co(u_S)$. Note that as $\edwa$ is copyless, for every $S$, any register $x
\in X$ appears in the expression $u_S$ at most once. Also note that
$\rho'$ may not be copyless as the registers $x_R$ denoting the same
subset $R$ may appear in two or more expressions $\rho'(q,a,x_S)$. The
output function can be define in a similar fashion by defining the
expression $u_S$ to be $\sum_{x\in S} \mu(q)$, and setting $\mu'(q) =
x_{reg(u_S)} + co(u_S)$.
\qed

\begin{figure}[t]
\centering
\begin{tikzpicture}[->,-stealth',semithick]

\node[state] (q0c) {$q_0$};
\node[coordinate,above left of=q0c,node distance=7mm] (start0c) {};
\draw (start0c) to (q0c);

\draw (q0c) to[loop above] node[smalltext,above]
          {$a\left/\hspace{-.5em}\begin{array}{l} x:=x \\
                                                  y:=y \sqp 1 \\
                                                  z:= z
                                 \end{array}
             \right.$}
      (q0c);

\draw (q0c) to[loop below] node[smalltext,below]
          {$b\left/\hspace{-.5em}\begin{array}{l} x:=x \\
                                                  y:=y \\
                                                  z:= z \sqp 1 \\
                                 \end{array}
             \right.$}
      (q0c);

\draw (q0c) to[loop right] node[smalltext,right]
          {$e\left/\hspace{-.5em}\begin{array}{l} x:=x \sqp y \sqp z\\
                                                  y:=0 \\
                                                  z:=0 \\
                                 \end{array}
             \right.$}
      (q0c);
\node[smalltext,below of=q0c,node distance=35mm] {$\mu(q_0) = x\sqp y\sqp z$};
\node[captiontext,below of=q0c,node distance=45mm] {(a) \EDWA$(+)$};

\node[state,node distance=50mm,right of=q0c] (q0) {$q_0$};
\node[coordinate,above left of=q0,node distance=7mm] (start0) {};
\draw (start0) to (q0);
\draw (q0) to[loop above] node[smalltext,above] (t) {} (q0);

\node[smalltext,node distance=15mm,above of=t]
          {$a\left/\begin{array}{l@{\hspace{-.1em}}l@{\hspace{.2em}}l}
            x_{\setof{x}}     & := & x_{\setof{x}} \\
            x_{\setof{y}}     & := & x_{\setof{y}} \sqp 1 \\
            x_{\setof{z}}     & := & x_{\setof{z}} \\
            x_{\setof{x,y}}   & := & x_{\setof{x,y}} \sqp 1 \\
            x_{\setof{y,z}}   & := & x_{\setof{y,z}} \sqp 1 \\
            x_{\setof{x,z}}   & := & x_{\setof{x,z}} \\
            x_{\setof{x,y,z}} & := & x_{\setof{x,y,z}} \sqp 1
            \end{array}\right.$};

\draw (q0) to[loop below] node[smalltext,below] (u) {} (q0);

\node [coordinate,node distance=15mm,below of=u] (ub) {};
\node [smalltext,right of=ub,node distance=5mm]
          {$b\left/\begin{array}{l@{\hspace{-.1em}}l@{\hspace{.2em}}l}
            x_{\setof{x}}     & := & x_{\setof{x}} \\
            x_{\setof{y}}     & := & x_{\setof{y}} \\
            x_{\setof{z}}     & := & x_{\setof{z}} \sqp 1 \\
            x_{\setof{x,y}}   & := & x_{\setof{x,y}} \\
            x_{\setof{y,z}}   & := & x_{\setof{y,z}} \sqp 1 \\
            x_{\setof{x,z}}   & := & x_{\setof{x,z}} \sqp 1 \\
            x_{\setof{x,y,z}} & := & x_{\setof{x,y,z}} \sqp 1
           \end{array}\right.$};

\draw (q0) to[loop right] node[smalltext,right]
          {$e\left/\hspace{-.5em}
            \begin{array}{l@{\hspace{-.1em}}l@{\hspace{.2em}}l}
                x_{\setof{x}}     & := & x_{\setof{x,y,z}}\\
                x_{\setof{y}}     & := & 0  \\
                x_{\setof{z}}     & := & 0 \\
                x_{\setof{x,y}}   & := & x_{\setof{x,y,z}} \\
                x_{\setof{y,z}}   & := & 0 \\
                x_{\setof{x,z}}   & := & x_{\setof{x,y,z}}\\
                x_{\setof{x,y,z}} & := & x_{\setof{x,y,z}}\\
            \end{array}\right.$} (q0);
\node[smalltext,node distance=35mm,below of=q0] {$\mu(q_0) = x_{x,y}$};
\node[captiontext,below of=q0,node distance=45mm] {(b) Corresponding \EDWA$(+c)$};

\end{tikzpicture}
\caption{Translation from copyless \EDWA$(+)$ to \EDWA$(+c)$\label{subsetc}}
\end{figure}
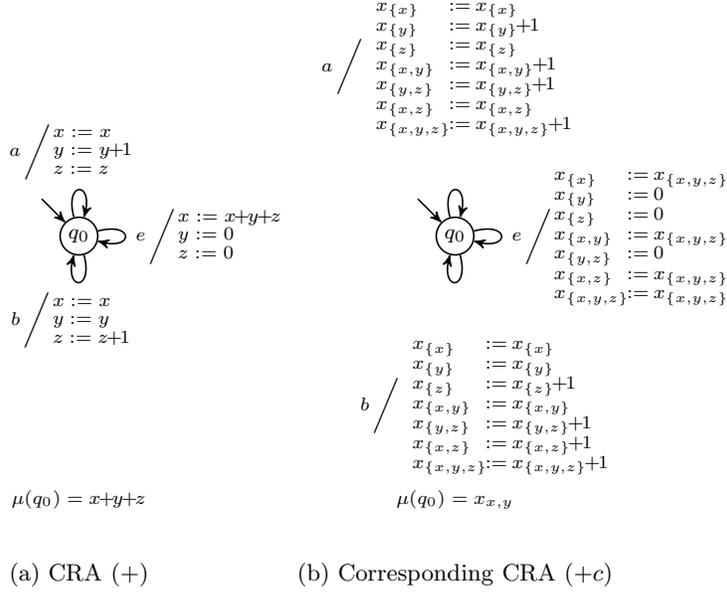

See Fig. \ref{subsetc} for an example of this construction.  The CRA on
the right of the figure will have one register for every possible subset
of registers of the CRA on the left of the figure. Let's consider the
transition on the symbol $b$.  In the original CRA the update performed on
$x$ is $x:=x+y+z$ and all the other registers are reset to $0$.  This
means that for all the $S$ containing $x$, $x_S$ is updated to
$x_{\setof{x,y,z}}$ while all the other registers are reset.

\begin{lemma}
\label{ssttopcedwa}
For a cost domain $\domain$ with a commutative and associative operation
$\interp{\myplus}$, $\reg{\domain,\myplus c}\subseteq
\CCF(\domain,\myplus)$.
\end{lemma}

\Proof~ Consider a function $f:\Sigma\st\mapsto\domain$ belonging to
$\reg{\domain,\myplus c}$, \ie, there is an \SSTT
$\stt=(Q,q_0,X,\delta,\rho,\mu)$ over the cost grammar $\CG(\myplus c)$
such that $\interp{\stt} = f$, for the cost model
$(\domain,\myplus c,\interp{.})$.  We show how we can construct a
copyless \EDWA~
$\edwa=(\inputalph,\edwastates,\edwainitst,\edwavariables,\edwatrans,\edwavarup',\edwafinal')$
over $(\domain,\myplus)$ that also computes $f$.

The \EDWA faithfully mimics the computation of the \SSTT $\stt$ in its
state. The only difference is the register update function and the
final output function.  The translation ensures that $\edwa$ maintains
the invariant that if a configuration of $\stt$ is
$(\edwastate,\valuation)$, the corresponding configuration for $M$ is
$(\edwastate,\valuation')$ such that for all $x\in X$,
$\interp{\valuation(x[0])} = \valuation'(x)$.

A register update expression $\rho(q,a,x)$ for $\stt$ has one of the
following forms: $x:=\myplus(x,c)$, $x:=c$, $\{x:=x[y],y:=\hole\}$.
Except for the last assignment, each RHS expression is in
$E(\CG(\myplus c),X)$, and with $\interp{\myplus} = +$, can be can be
directly mimicked by $\edwa$ by setting $\rho'(q,a,x)$ to be the
expressions  $x:=x+c$ and $x:=c$ respectively.  To simulate
$\{x:=x[y],y:=\hole\}$, we note that as $\myplus$ is associative and
commutative, for a term in $E(\CG(\myplus c),X)$, the term $x[y]$ is
equivalent to the term $\myplus(x,y)$. Thus, with $\interp{\myplus} =
+$, $\edwa$ can simulate parameter substitution by the {\em copyless}
assignment $\{x:=x\myplus y, y:=0\}$.  The output function $\mu'(q)$ can be
mimicked in a similar fashion for the corresponding expressions in
$\mu(q)$.
\qed

\begin{theorem}[Expressiveness of Additive Cost Functions]\label{regplus}
For cost domain $\domain$ with a commutative associative operation $\myplus$,
$\CF(\domain,\myplus c)=\CCF(\domain,\myplus)=\reg{\domain,\myplus c}=\reg{\domain,\myplus}$.
\end{theorem}
\Proof~ Follows from Lemma
\ref{ssttostt},\ref{mvcmrtosst},\ref{ppcedwatopedwa} and
\ref{ssttopcedwa}.
\qed \\

We can also establish the following results establishing an
expressiveness hierarchy between different classes.  First we show
that the copyless restriction for \EDWA over $(\domain, \myplus c)$
reduces the expressivity.

\begin{theorem}\label{thm:plus-c-copying-essential}
$\CF(\domain,\myplus c)\nsubseteq\CCF(\domain,\myplus c)$.
\end{theorem}

\Proof~ Consider the function $f_1$ computed by the \EDWA $M_1$ in
Fig.  \ref{rcfex1} that maps a string $w$ to the length of the
substring obtained by deleting all $b$'s after the last occurrence of
$e$ in $w$. We show that for any fixed $k$, there does not exist a
copyless \EDWA capable of computing this function.

Assume that \edwa~is a copyless \EDWA~ over $(\domain,\myplus c)$ that
can compute this function with $k$ registers.  Without loss of
generality we can assume that in every assignment of the form $x:=
y+c$, $y$ and $x$ are the same register (if the original machine is
doing some copyless renaming, we can remember the renaming in the
state).

Now consider a string of the form $w=b^{n}aeb^{n}ae\ldots$, where
$n$ is greater than the number of states in $M$. Thus, for each
$i$, while processing the $i^{\mbox{th}}$ block of $b^{n}$, some
state $q_{i}$ (possibly depending on the block number $i$) must
be visited at least twice. For each $i$, let $x_{i}$ be the register
used to calculate the output after reading the input string $\left(b^{n}ae\right)^{i}a$.
Now observe that for all $i<j$, $x_{i}$ and $x_{j}$ are necessarily
distinct: if we pump $w$ to $w^{\prime}=\left(b^{n}ae\right)^{i-1}b^{n+l}ae\left(b^{n}ae\right)^{j-i}\ldots$,
the value of $x_{i}$ after $i$ blocks must be unchanged, but the
value of $x_{j}$ must change. We have thus established that a different
register must be used to produce the output after each block, but
this is not possible if the machine has only a finite number ($k$)
of registers.
\qed

We then show that removing the copyless restriction from CRAs over
$(\domain, \myplus)$ is too permissive as it allows computing cost
functions that grow exponentially.

\begin{theorem}
\label{mvcmtomvcmr}
There exists a \EDWA~\edwa~over $(\domain,\myplus)$ that cannot be
expressed as a copyless \EDWA~\edwa~over $(\domain,\myplus)$, \ie,
$\CF(\domain,\myplus)\nsubseteq\CCF(\domain,\myplus)$.
\end{theorem}

\Proof~We prove this result by contradiction.  We create a
\EDWA~\edwa~over $(\domain,\myplus)$ over the alphabet $\{a\}$ such
that on the first $a$ it perform the update $x:=2$ and on the
subsequent $a$'s it performs the update $x:=x+x$. Given a string $w\in
a^+$, the function computed is $f=2^{|w|}$.  Let's assume that there
exists a copyless \EDWA~$\edwa'$ that can compute $f$.  By Theorem
\ref{regplus}, there must be an \SSTT~$\stt$ in $\reg{\domain, \myplus
c}$ that also computes $f$.  However, as the output is not linearly
bounded, this function cannot be computed by an \SSTT and so we reach
a contradiction.  \qed \\

Finally we show that having multiple registers is essential for
expressive completeness.

\begin{theorem}\label{kvar}
For every $k\in\mathbb{N}$, there is a cost function $f$ so that every
\EDWA~$M$ over $\CF(\Nat,+c)$ has at least $k$ registers.
\end{theorem}

\Proof~For each $k$, consider the function
$f_{k}:\mathbb{N}\to\mathbb{N}$ defined as
$f_{k}\left(x\right)=\left((x\bmod k)+1\right).x$.  The input $x$ is
expressed in a unary alphabet $\Sigma=\setof{1} $.  This function
outputs one of $x$, $2x$, $3x$, \ldots{}, $kx$, depending on the
length of $x$.

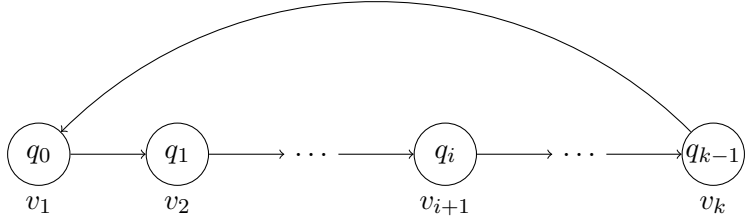
\begin{figure}
\begin{centering}
\begin{tikzpicture}

  \node [place] (q0)    [              label=below:$v_{1}$  ] {$q_{0}$};
  \node [place] (q1)    [right=of q0,  label=below:$v_{2}$  ] {$q_{1}$};
  \node         (el1)   [right=of q1                        ] {$\ldots$};
  \node [place] (qi)    [right=of el1, label=below:$v_{i+1}$] {$q_{i}$};
  \node         (el2)   [right=of qi                        ] {$\ldots$};
  \node [place] (qkm1)  [right=of el2, label=below:$v_{k}$  ] {$q_{k-1}$};

  \draw [->] (q0)   to                 (q1);
  \draw [->] (q1)   to                 (el1);
  \draw [->] (el1)  to                 (qi);
  \draw [->] (qi)   to                 (el2);
  \draw [->] (el2)  to                 (qkm1);
  \draw [->] (qkm1) to [out=135, in=45] (q0);

\end{tikzpicture}
\par\end{centering}

\caption{\label{fig:k-var-reqd:MRAA:M_k}\EDWA~ $M_{k}$ implementing $f_{k}$.
Each transition performs $v_{i}:=v_{i}+i$, for all $i$.}
\end{figure}

First, these functions can be implemented by a \EDWA $\edwa_k$ (shown
in \Fig \ref{fig:k-var-reqd:MRAA:M_k}).  $\edwa_k$ has $k$ registers
$v_{1}$, \ldots{}, $v_{k}$, all initialized to $0$ and has $k$ states
$q_{0}$, \ldots{}, $q_{k-1}$.  For each $i$, $\delta(q_i,1) =
q_{(i+1)\bmod k}$, $\rho(q_i,1,v_i) = v_i + i$, and $\mu(q_i) =
v_{(i+1)\bmod k}$.

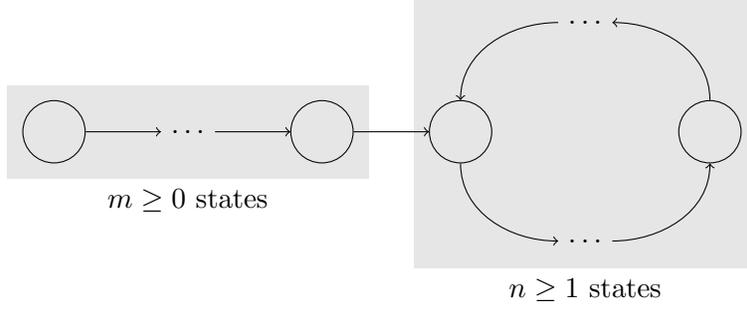
\begin{figure}
\begin{centering}
\begin{tikzpicture}

  \node [place] (q0)                       {};
  \node         (el1) [right=of q0]        {$\ldots$};
  \node [place] (q1)  [right=of el1]       {};

  \node [place] (q2)  [right=of q1]        {};
  \node         (el2) [below right=of q2]  {$\ldots$};
  \node [place] (q3)  [above right=of el2] {};
  \node         (el3) [above left=of q3]   {$\ldots$};

  \draw [->] (q0)  to                   (el1);
  \draw [->] (el1) to                   (q1);
  \draw [->] (q1)  to                   (q2);

  \draw [->] (q2)  to [out=270, in=180] (el2);
  \draw [->] (el2) to [out=0, in=270]   (q3);
  \draw [->] (q3)  to [out=90, in=0]    (el3);
  \draw [->] (el3) to [out=180, in=90]  (q2);

  \begin{pgfonlayer}{background}

    \node [background, fit=(q0) (el1) (q1), label=below:$m \geq 0$ states] {};
    \node [background, fit=(q2) (el2) (q3) (el3), label=below:$n \geq 1$ states] {};

  \end{pgfonlayer}

\end{tikzpicture}
\par\end{centering}
\caption{\label{fig:k-var-reqd:MRAA:Mkm1:Structure}The structure of a possible
$k-1$-register \EDWA~ $M$ implementing $f_{k}$.}
\end{figure}

We now show that at least $k$ registers are necessary. Consider
otherwise, and say we are able to produce such an \EDWA~ $M$ with
$k-1$ registers.  The main idea is that the difference between any two
{}``sub''-functions $ax$ and $bx$, $a\neq b$, grows without bound
Since $M$ works over a unary input
alphabet, the only form it can assume is that of a lasso (\Fig
\ref{fig:k-var-reqd:MRAA:Mkm1:Structure}). Say there are $m\geq0$
states in the initial approach to the loop, and $n\geq1$ states in a
single pass of the loop.

In the following argument, let $c$ denote the largest constant that
appears in the description of $M$.  Without loss of generality, we can
assume that no register renaming occurs, as for a \EDWA with register
renaming, there is an equivalent \EDWA with no register renaming, by
tracking register renaming as part of its state. Thus, if in state
$q$, if $\rho(q,1,v_i)$ is the expression $v_j + c_1$ (where $i\neq
j$), then there is no $v_\ell$ such that $v_i$ appears in
$\rho(q,1,v_\ell)$.

Also observe that no register that ever gets reset during the loop can
contribute to the output afterwards. Say there is some transition
during which $v_{i}$ is updated as $v_{i} := c_{3}$. If this register
influences the output $n^{\prime}\geq0$ states later, then for $x > c
n^{\prime} + c_{3}$, the output must be incorrect.

Now pick some state $q$ occurring in the loop. Say that the output in
$q$ depends on some register $v_{i}$. Let us call $q$ good if register
$v_{i}$ in $q$ influences the output in some state $q^{\prime}$ which
is $p < k$ transitions from $q$. At least one good state has to exist,
since the machine has at most $k-1$ registers. We now use the presence
of this good state to derive a contradiction: the outputs in $q$ and
$q^{\prime}$ can differ by no more than $(p + 2).c$. But since $q$ and
$q^{\prime}$ are closer than $k$ steps apart, they output necessarily
different functions, and hence for $x > (p + 2).c$, the outputs in
these two states are required to differ by more than this amount. The
contradiction is complete.
\qed \\

We now prove some closure properties of the model.

\begin{theorem}[Addition]
Given two CRAs $\edwa_1$ and $\edwa_2$ over the cost model $(\Rat,+)$,
there exists a \edwa~over the same cost model such that $\forall
w\in\Sigma\st.$ $M(w)=M_1(w)+M_2(w)$.
\end{theorem}

\Proof~ Given $\edwa_i = (\inputalph, \edwastates^i, \edwainitst^i,
\edwavariables^i, \edwatrans^i, \edwavarup^i, \edwafinal^i)$ (where
$i\in\{1,2\}$) we construct $\edwa=(\inputalph,
\edwastates^1\times\edwastates^2, (\edwainitst^1,\edwainitst^2),
\edwavariables^1\times\edwavariables^2\cup\edwavariables^1\cup\edwavariables^2,
\edwatrans, \edwavarup, \edwafinal)$. We assume $\edwavariables_1$ and
$\edwavariables_2$ are disjoints sets.  The registers in $\edwa$ are
pairs of the form $(x,y)$ or singletons of the form $x$.  Whenever
$\edwa_i$, while processing an input word $w$, is in the configuration
$(\edwastate_i,\valuation_i)$, $\edwa$ has the configuration
$((\edwastate_1,\edwastate_2),\valuation)$ such that:
(1) $\valuation((x,y))=\valuation_1(x)+\valuation_2(y)$,
(2) $\valuation(x)=\valuation_1(x)$ if $x\in\edwavariables^1$, and
(3) $\valuation(y)=\valuation_2(y)$ if $y\in\edwavariables^2$.

We now define the update functions $\edwatrans$ and $\edwavarup$.
Given $(v_1,v_2)\in\edwastates_1\times\edwastates_2, a\in\inputalph,
(x_1,x_2)\in\edwavariables^1\times\edwavariables^2$, define
$\edwatrans((v_1,v_2),a)=(\edwatrans^1(v_1,a),\edwatrans^2(v_2,a))$,
and
$\edwavarup((v_1,v_2),a,(x_1,x_2))=reg(\edwavarup^1(v_1,a,x_1)+\edwavarup^2(v_1,a,x_2))+co(\edwavarup^1(v_1,a,x_1)+\edwavarup^2(v_1,a,x_2))$.

Given an expression $e$, $reg(e)$ is equal $x$
when $x$ is the only variable appearing in $e$, to $(x,y)$ when
$x\in\edwavariables^1,y\in\edwavariables^2$ are the only two variable
appearing in $e$ and $\varepsilon$ when no variable appears in $e$,
while $co(e)$ is the sum of all the constants appearing in $e$.  Note
that, at every point, at most 2 registers can appear in the combined
right-hand side.  The output function can be defined in a similar way.
\jyo{We need a little more explanation here.} By construction $\edwa$
computes the right function.
\qed

\begin{theorem}[Subtraction]\label{subtraction}
Given two CRAs $\edwa_1$ and $\edwa_2$ over the cost model $(\Rat,+)$,
there exists a $\edwa$ over the same cost model such that
$\forall w\in\Sigma\st.$ $M(w)=M_1(w)-M_2(w)$.
\end{theorem}
\Proof~ Similar to the proof for addition.
\qed \\

\mypar{Weighted Automata}\label{wa}
A weighted automaton~\cite{droste_handbook_2009} over an input
alphabet $\Sigma$ and a cost domain $\domain$ is a {\em
nondeterministic\/} finite-state automaton whose edges are labeled
with input symbols in $\Sigma$ and costs in $\domain$.  For an input
string $w$, the automaton can have multiple accepting paths from its
initial state to an accepting state.  The semantics of the automaton
is defined using two binary functions $\mytimes$ and $\myplus$ such
that $\mytimes$ is associative and commutative, and $\myplus$
distributes over $\mytimes$ (to be precise, form a {\em semiring\/}
algebraic structure).  The cost of a path is the sum of the costs of
all the transitions along the path according to $\myplus$, and the
cost of a string $w$ is obtained by applying $\mytimes$ to the set of
costs of all accepting paths of the automaton over $w$.

Let $K$ be the semiring $(\srset,\mytimes,\myplus,\srzero, \srone)$.
Formally, a weighted automaton \WA with weights from $K$, from an
input alphabet $\Sigma$ into the domain $\srset$ is a tuple $\wa =
(\inputalph, \wastates, \wainitst, \wafinalst, \watrans, \wainit,
\wafinal)$ where $\Sigma$ is a finite input alphabet, $\wastates$ is a
finite set of states, $\wainitst\subseteq\wastates$ the set of initial
states, $\wafinalst\subseteq\wastates$ the set of final states,
$\watrans$  a finite multiset of transitions, which are elements of
$\wastates\times \Sigma \times \srset\times \wastates$, $\wainit :
\wainitst \mapsto \srset$ an initial weight function, and $\wafinal :
\wafinalst \mapsto \srset$ a final weight function mapping
$\wafinalst$ to $\srset$.

Consider a string $w=w_1\ldots w_n \in\Sigma\st$ and a \WA~\wa.  A
sequence $\pi=(q_0,c_0),(q_1,c_1),\ldots, (q_n,c_n)$ is an accepting
in sequence for $w$ if, for every $1\leq i\leq n$,
$(q_{i-1},w_i,c_i,q_i)\in\watrans$ and $\wainit(q_0)=c_0$.  The weight
of $\pi$ (denoted as $w(\pi)$) is computed as $(\myplus_{0\leq i\leq
n}c_i) \myplus\wafinal(q_n)$.  Given a word $w$ we denote by $P(w)$
the set of all the accepting paths of $w$.  The weight $T(w)$ of the
string $w$ is defined as: $$T(w)=\bigoplus_{\wapath\in P(s)} w[\pi]$$
where $\srplus$ is also an operation over $S$.  A weighted automaton
is called {\em single-valued} if each input string has at most one
accepting path\footnote{In some of the literature these are called
\emph{unambiguous} weighted automata, while a \emph{single-valued}
weighted automaton is one where the weights of all the accepting paths
are the same. The two notions are proved to be equivalent.}.  To
interpret a single-valued weighted automaton, we need only an
interpretation for $\myplus$. Thus, we can compare the class of
functions definable by such automata with regular additive functions.

\begin{theorem}[Single-valued Weighted Automata]\label{exprplus}
A cost function $f:\fm\Sigma\mapsto\domain$ is in $\reg{\domain,\myplus c}$
iff it is definable by a single valued weighted automaton.
\end{theorem}
\Proof~ Let $\wa=(\inputalph, \wastates, \wainitst, \wafinalst,\watrans,
\wainit, \wafinal)$ a \emph{single valued} weighted automaton that
computes the function $f:\fm\Sigma\mapsto\domain$.  We construct an
\SSTT with \emph{regular look-ahead}
$(\stt,A)=(\wastates\cup\{\wainitst\},\wainitst,\{v\},\delta,\rho,\mu)$,
$(R,r_0,\delta_R)$ that computes $f$. The \SSTT $\stt$ uses the cost
grammar $\CG(\mytimes c)$ to construct its terms.  We then use
\thmref{stt-rla} to show that there exists an \SSTT~ that computes
$f$, which means that $f$ is regular.

Even though $\wa$ is nondeterministic, since it is \emph{single
valued}, it will have only one accepting path.  We construct
$A=(R,r_0,\delta_R)$ such that the states in $R$ give information on
what is the next transition to take to reach an accepting path.  Every
state $r\in R$ is a pair $(a,f)$ where
$a\in(\Sigma\cup\setof{\varepsilon})$ and $f$ is a partial function
from $\wastates$ to $\wastates$. After reading the $i^{th}$ symbol of
the input word $w=w_1\ldots w_n$, $r=(w_i,f)$ and $f(q)=q'$ if:
1) $(q,a,c,q')\in \watrans$ for some $c$, and
2) if $\wa$ starts reading $w_{i+1}\ldots w_n$ in $q'$, it will reach an accepting state.

The initial state $r_0$ is defined as $(\varepsilon,f_0)$.  and for
every $q\in\wafinalst$ $f_0(q)=q$. The initial state does not encode
any information as it corresponds to the case where the $\stt$ has
reached the end of the string $w$.  We now define $\delta_R$. Suppose
$A$ is in state $(a,f)$ and it is reading the input $b$. The new state
will be $(b,f')$ where $f'(q)=q'$ if $f(q')$ is defined and
$(q,b,c,q')\in \watrans$ for some $c$.

We now define the state transition function for the \SSTT $\stt$. We
define $\delta(q,(a,f))=q'$ if $f(q)=q'$.  Particular attention must
be made for the case when $\stt$ is in state $\wainitst$.  In this
case on input symbol $(a,f)$, $\delta(\wainitst,(a,f))=q'$ where $q'$
is the only state in $\wainitst$ such that $f(q)=q'$. Notice that
there can be only one state of this form otherwise $\wa$ would not be
single valued.  For every transition of the form $(q,a,c,f(q))$ in
$\wa$, in $\stt$, the register update function $\rho(q,a,v)$ maps $v$
to the expression $\myplus(v,c)$. This shows that for every weighted
automaton $\wa$, we can construct an \SSTT $\stt$ over the cost
grammar $\CG(\mytimes c)$ such that for all input strings $w$,
$\interp{\stt}(w) = \wa(w)$.

We now prove the other direction.  By Theorem~\ref{regplus} we know
that every function $f$ in $\reg{\domain,\myplus c}$ can be computed
by a \EDWA $\edwa$ over $(\domain, \myplus c)$. Let
$\edwa=(\inputalph,\edwastates,\edwainitst,\edwavariables,\edwatrans,\edwavarup,\edwafinal)$.
We show how we can construct a \emph{single valued} weighted automaton
$\wa=(\inputalph, \edwastates\times(\edwavariables\cup\{r\}),
\{\edwainitst\}\times(\edwavariables\cup\{r\}),
\edwastates\times\edwavariables,\watrans, \wainit, \wafinal)$ that
also computes $f$.

After processing an input word $w$, if $\edwa$ has the configuration
$(\edwastate,\valuation)$, we have that: corresponding to every $x\in
X$ there exists a path in $\wa$ from the initial state such that the
cost along that path is equal to $\valuation(x)$.  Let's now give the
definition of $E$:
(1) if $\delta(q,a)=q'$ and $\rho(q,a,x)=\myplus(y,c)$, then $((q,y),a,c,(q',x))\in E$,
(2) if $\delta(q,a)=q'$ and $\rho(q,a,x)=c$, then $((q,r),a,c,(q',x))\in E$, and
(3) if $\delta(q,a)=q'$, $((q,r),a,0,(q',r))\in E$,
The nodes $(q,r)$ are always reachable with cost $0$ and are used to
represent resets, but none of them is accepting. $\wafinal$ can be
defined in a similar way.

A simple inductive proof establishes that in a \EDWA over
$(\domain,\myplus c)$, in any state, only one register eventually
contributes to the final output, or in other words, only one value
flows to the final output. Thus, the constructed weighted automaton is
is single-valued.
\qed

\begin{figure}[t]
\centering
\begin{tikzpicture}[->,-stealth',semithick]

\node[state] (q0) {$(q_0,x)$};
\node[state,right of=q0,node distance=30mm] (q1) {$(q_0,y)$};
\node[coordinate,above left of=q0,node distance=10mm] (start0) {};
\node[coordinate,above left of=q1,node distance=10mm] (start1) {};

\node[smalltext,above left of=start0,node distance=2mm] {$0$};
\node[smalltext,above left of=start1,node distance=2mm] {$0$};

\draw (start0) to (q0);
\draw (start1) to (q1);

\draw (q1) to node[smalltext,above] (mid) {$e, +1$} (q0);
\draw (q0) to[loop above] node[smalltext,above] {$a, +1$} (q0);
\draw (q0) to[loop below] node[smalltext,below] {$b, 0$}  (q0);
\draw (q1) to[loop above] node[smalltext,above] {$a, +1$} (q1);
\draw (q1) to[loop below] node[smalltext,below] {$b, +1$} (q1);
\draw (q1) to[loop right] node[smalltext,right] {$e, +1$} (q1);

\node[smalltext,node distance=17mm,below of=mid] {$Out((q_0,x)) = 0$};
\node[smalltext,node distance=20mm,below of=mid] {$Out((q_0,y)) = \infty$};

\end{tikzpicture}
\caption{Weighted Automaton corresponding to $M_1$ in Figure~\ref{rcfex1}\label{wam1}}
\end{figure}
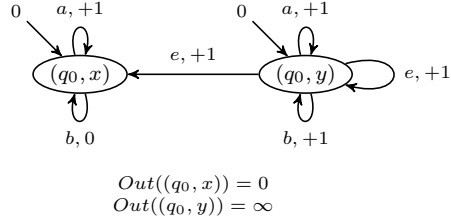

An example of the translation of the function $M_1$ of
Figure~\ref{rcfex1} is in Figure~\ref{wam1}. The machine has only two
states $(q_0,x)$, and $(q_0,y)$. If we take for example the transition
of the first automaton when reading $c$, we can see that $x$ is
updated to $y+1$. In the automaton of Figure~\ref{wam1} this is
reflected by the transition from $(q_0,y)$ to $(q_0,x)$ with label $c$
and with weight $+1$.

\mypar{Decision Problems}
\paragraph{Minimum Costs.}
The shortest path problem for CRAs is to
find a string $w$ whose cost is the minimum.
For numerical domain with addition, for CRAs with increment, we can solve the shortest path problem by reducing
it to classical shortest paths using the translation from CRAs with increment to single-valued weighted automata
used in the proof of Theorem~\ref{exprplus}.
If the CRA has $n$ states and $k$ registers, the graph has $n\cdot k$ vertices.
The exact complexity depends on the weights used: for example, if
the costs are nonnegative, we can use Dijkstra's algorithm.

\begin{theorem}[Shortest Path for CRAs with Inc]\label{pEDWAmcp}
Given a CRA $\edwa$ over the cost model $(\Rat,+c)$, computing $\min\{\edwa(w)\sep w\in\fm\Sigma\}$
is solvable in $\ptime$.
\end{theorem}

\Proof~ We reduce the problem to shortest finding the shortest path in
a weighted graph. Using the construction of Theorem~\ref{exprplus} we
create a weighted graph. The graph has $nk$ nodes and
$|\Sigma|nk$ edges where $n,k$ are the number of states and variable of $\edwa$ respectively.  If the weights are all positive we can use
Dijkstra's algorithm with a final complexity of
$\bigO(|\Sigma|nk+nk \log(nk))$ otherwise we can use
the Bellman-Ford algorithm, making the complexity $\bigO(n^2k^2
\log(nk))$.  \qed \\

\begin{figure}[t]
\centering
\begin{tikzpicture}[->,-stealth',semithick]

\node[state] (q0) {$q_0$};
\node[coordinate,left of=q0,node distance=7mm] (start) {};
\draw (start) to (q0);

\draw (q0) to[loop above] node[smalltext,above]
     {$b\left/\hspace{-.5em}\begin{array}{l} x := x \\
                                          y := y \sqp 1 \\
                            \end{array}\right.$}
      (q0);

\node[coordinate,right of=q0,node distance=40mm] (mid) {};

\node[state,node distance=20mm,above of=mid] (q1) {$q_1$};
\node[state,node distance=20mm,below of=mid] (q2) {$q_2$};

\draw (q1) to[loop above] node[smalltext,above]
     {$b\left/\hspace{-.5em}\begin{array}{l} x:= x \\
                                              y:=y\sqp 1 \\
                             \end{array}\right.$}
      (q1);

\draw (q2) to[loop below] node[smalltext,below]
     {$b\left/\hspace{-.5em}\begin{array}{l} x:= x \\
                                              y:=y\sqp 1 \\
                             \end{array}\right.$}
      (q2);

\draw (q0) to node[smalltext,sloped,above]
     {$a\left/\hspace{-.5em}\begin{array}{l} x:= x \sqp 1 \\
                                              y:= y \\
                             \end{array}\right.$}
      (q1);

\draw (q2) to node[smalltext,sloped,above]
     {$a\left/\hspace{-.5em}\begin{array}{l} x:= x \sqp 1 \\
                                              y:= y \\
                             \end{array}\right.$}
      (q0);

\draw (q1) to node[smalltext,right]
     {$a\left/\hspace{-.5em}\begin{array}{l} x:= x \sqp 1 \\
                                              y:= y \\
                             \end{array}\right.$}
      (q2);

\node[coordinate,node distance=20mm,right of=q0] (c) {};
\node[smalltext,below of=c,node distance=25mm]
        {$\begin{array}{l}
          \mu(q_0) = x \sqp y \\
          \mu(q_1) = x \\
          \mu(q_2) = y \\
          \end{array}$};
\end{tikzpicture}
\caption{Example CRA over $(\myplus)$ needing less registers than CRA
over $(\myplus c)$\label{succint}}
\end{figure}

Even though $\CF(\domain,\myplus c)=\CCF(\domain,\myplus)$, the model
with addition can be more succinct (see Fig. \ref{succint} for an
example).  To solve minimum-cost problem for copyless-CRAs over  the
cost model $(\domain,\myplus )$, we can use the translation to CRAs
over $(\domain,\myplus c)$ used in the proof of Theorem~\ref{regplus},
which causes a blow-up exponential in the number of registers. We can
establish an NP-hardness bound for the min-cost problem by a simple
reduction from 3-SAT.

\begin{theorem}[Shortest Paths for CRAs with Addition]\label{ppCEDWAmcp}
Given a copyless-CRA $\edwa$ over the cost model $(\Rat,+)$ with $n$
states and $k$ registers, computing $\min\{\edwa(w)\sep
w\in\fm\Sigma\}$ is solvable in time polynomial in $n$ and exponential
in $k$.  Given a copyless CRA $\edwa$ over the cost model $(\Nat,+)$ and a
constant $K\in\Nat$, deciding whether there exists a string $w$ such
that $\edwa(w)\leq K$ is \nphard.
\end{theorem}

\Proof~The first result follows from the complexity of the translation
in Lemma \ref{ppcedwatopedwa}.  For the second part we give a
reduction from 3-SAT.  Given an instance $V=\{v_1,\ldots,v_n\},
C=\{c_1,\ldots,c_k\}$ where $V$ is the set of literals and $C$ the set
of clauses we construct a \EDWA $\edwa$ over $(\Nat,+)$. $\edwa$ is
defined as the tuple ($\Sigma, Q, q_0, X, \delta, \rho, \mu$), where
$\Sigma = \setof{0,1}$, $Q = \setof{q_0,\ldots,q_{n+2}}$, $X =
\setof{x_1,\ldots,x_k}$. The update functions are defined as follows:
For each $i$, and $b \in \Sigma$, $\delta(q_i,b) = q_{i+1}$, and
$\rho(q_i,0,x_j) = 0$ if the clause $c_j$ becomes $\mathit{true}$ when
the variable $v_i$ is $\mathit{false}$.  Similarly $\rho(q_i,1,x_j) =
0$ if the clause $c_j$ becomes $\mathit{true}$ when $v_i$ is
$\mathit{true}$.  Finally, we define $\rho(q_{n+1},b,x_1) =
\sum_{i=1}^k x_i$, and $\mu(q_{n+2}) = x_1$.  It is easy to see that
every path in the $\edwa$ corresponds to a unique valuation for the
literals $v_1,\ldots,v_n$. Finally, if the minimum-cost computed by
$\edwa$ is $0$, then we have an instance of SAT, as there is a
valuation of the literals that makes every clause $\mathit{true}$. If
the minimum-cost computed is greater than $0$, then the conjunction of
the clauses is unsatisfiable.  This means that solving min-value
problem for a \EDWA over $(\Nat,+)$ is as hard as solving 3-SAT.
\qed \\

\paragraph{Equivalence and Containment.} Given two cost register
automata using addition over a numerical domain, checking whether they
define exactly the same function is solvable in polynomial time
relying on properties of systems of linear equations.

\begin{theorem}[Equivalence of CRAs with Addition]\label{thm:equiv-CRA-add}
Given two CRAs $\edwa_1$ and $\edwa_2$ over the cost model $(\Rat,+)$,
deciding whether for all $w$, $\edwa_1(w)=\edwa_2(w)$ is solvable in $\ptime$.
\end{theorem}
\Proof~
Given $\edwa_i = (\inputalph, \edwastates^i, \edwainitst^i,
\edwavariables^i, \edwatrans^i, \edwavarup^i, \edwafinal^i)$ (where
$i\in\{1,2\}$) we construct $\edwa=(\inputalph,
\edwastates^1\times\edwastates^2, (\edwainitst^1,\edwainitst^2),
\edwavariables^1\cup\edwavariables^2, \edwatrans, \edwavarup,
\edwafinal)$. We assume the two sets variables $\edwavariables_1$ and
$\edwavariables_2$ are disjoint.

For every $(q_1,q_2)\in\edwastates^1\times\edwastates^2$ and
$a\in\Sigma$,
$\edwatrans((q_q,1_2),a)=(\edwatrans_1(q_1,a),\edwatrans_2(q_2,a))$.
$\edwa$ updates the two constituent sets of registers separately. For
$x_1\in\edwavariables_1$,
$\edwavarup((q_1,q_2),a,x_1)=\edwavarup_1(q_1,a,x_1)$, and for
$x_2\in\edwavariables_2$
$\edwavarup((q_1,q_2),a,x_1)=\edwavarup_2(q_2,a,x_2)$.

We want to check if along every path of the $\edwa$, and for every
state $(q_1,q_2)$, the equation $\edwafinal_1(q_1)=\edwafinal_2(q_2)$
holds.  We adapt the algorithm for checking validity of affine
relations over affine programs presented in \cite{olm_note_2004} to do
this. The algorithm in \cite{olm_note_2004}, checks the validity of
affine relations (equality constraints over linear combinations of
real-valued or rational-valued program variables and constants), over
affine graphs (graphs where each edge is labeled by an affine
assignment). We can cast the equivalence check for CRAs over $(\Rat,
\myplus)$ as  a subcase of this problem.

The algorithm propagates the
equation $(\edwafinal_1(q_1)=\edwafinal_2(q_2)$ backward along each
transition using the register update function: for an edge from $q$ to
$q'$ with some label $a$, every equation $e_1=e_2$ that must hold at
$v$ yields an equation $e_1'=e_2'$ that must hold at $u$, where the
expressions $e_1'$ and $e_2'$ are obtained from $e_1$ and $e_2$ using
substitution to account for the update of registers along the edge
from $u$ to $v$.  $e_i'$ will be equal to $e_i$ where every register
$x\in\edwavariables$ is replaced by $\edwavarup(u,a,x)$.  At every
step of the back propagation, we compute the basis of the set of
equations in every state using Gaussian elimination.  If we reach a
system of equations with no solution the two machines are
inequivalent, while if we reach a fix point where no independent
equations can be added, the two machines are equivalent.

As shown in Theorem 2 of \cite{olm_note_2004}, such a propagation
terminates in $\bigO(nk^3)$ where $n$ is the size of the machine (in
our case $|\edwastates^1||\edwastates^2||\inputalph|$) and $k$ is the
number of registers (in our case
$|\edwavariables^1|+|\edwavariables^2|$).


For CRAs that use only increment, the cubic complexity of the Gaussian
elimination in the inner loop of the equivalence check can be
simplified to quadratic: at every step  in the back propagation, all
equations are of the form $x=y+c$.  The final complexity is
$\bigO(nk^2)$
if we only have increments and
$\bigO(nk^3)$
otherwise ($k,n$ are as defined before).
\qed

We now show that general containment is also decidable in polynomial
time and that checking if a number is in the range of a CRA over
$(\Int,+)$ is decidable in polynomial time.

\begin{theorem}[$\edwa_1\leq\edwa_2$]
Given two CRAs $\edwa_1$ and $\edwa_2$ over the cost model $(\Rat,+)$,
deciding whether $\forall w\in\fm\inputalph. \edwa_1(w)\leq\edwa_2(w)$
is in $\ptime$.
\end{theorem}
\Proof~
We reduce the problem to shortest path. We in fact have that if
$\forall w\in\fm\inputalph.\edwa_1(w)\leq\edwa_2(w)$ then $\forall
w\in\fm\inputalph.\edwa_1(w)-\edwa_2(w)\geq 0$. But from Theorem
\ref{subtraction} we can construct an $\edwa$ which is equivalent to
$\edwa_1(w)-\edwa_2(w)$ and has polynomial size. Now we can solve
shortest path on $\edwa$. The algorithm is clearly polynomial.
\qed

\begin{theorem}[$k\in$ Range]
Given two CRAs $\edwa$  over the cost model $(\Int,+)$ and a constant $k\in\Int$,,
deciding whether $\exists w\in\fm\inputalph. \edwa(w)=k$ is in $\nlogspace$.
\end{theorem}
\Proof~
We reduce the problem to 0 reachability in a weighted graph. Using Theorem \ref{subtraction} we compute $\edwa'=\edwa-c$.
Now we want to check $\exists w\in\fm\inputalph. \edwa'(w)=0$. We can create the same graph of shortest path
and look for a 0 path on it. The problem of $0$-reachability over finite graphs is known to be in $\nlogspace$.
\qed

\section{Semiring Cost Models}\label{sec:scm}

In this section, we consider the cost models which result when the
cost model supports two binary operations,
$\sradd$ and $\srmul$, that impose a semiring structure (see subsection \ref{wa} for the definition of semiring).
This structure has been studied extensively in the literature on weighted automata and rational power series.
A specific case of interest is the {\em tropical semiring\/},
where the cost domain is $\Nat\cup \left\{\infty\right\}$, $\sradd$ is the $\min$ operation, and $\srmul$ is arithmetic addition.
While choosing a grammar, we can restrict either or both of $\sradd$ and $\srmul$ to be ``unary''
(that is, the second argument is a constant).
To study the tropical semiring, it makes sense to choose $\min$ to be binary, while
addition to be unary. Hence, in this section, we will focus on the grammar $\CG(\sradd,\srmul c)$,
and the class $\reg{\domain,\sradd,\srmul c}$ of cost functions.

\mypar{CRA Models}
Our first task is to find a suitable set of operations for cost register automata so as
to have expressiveness same as the class  $\reg{\domain,\sradd,\srmul c}$.
It turns out that (unrestricted) CRAs with $\sradd$ and $\srmul c$ are too expressive,
while their copyless counterparts are too restrictive.
We need to enforce the copyless restriction, but allow substitution.
In the proposed model, each register $x$
has two fields ranging over values from $\domain$: $\left(x.c,x.d\right)$.
The intuitive understanding is that $x$ represents the expression
$\left(x.d\, \srmul\, \hole \right)\, \sradd\, x.c$ where $\hole$ denotes the parameter.
Such a pair can be viewed as the ``most evaluated'' form
of a parameterized term in the corresponding \SSTT.
Expressions used for the update are given by the grammar
\begin{align*}
e & ::=\left(c,d\right)\mid x\mid e_{1} \dadd e_{2}\mid e_{1} \dscale d\mid e_{1}\left[e_{2}\right]
\end{align*}
where $x$ is a register, and $c$ and $d$ are constants.
For the min-inc interpretation, the initial values are of the form
$\left(\infty,0\right)$ corresponding to the additive
and multiplicative identities.
We require that registers are used in a copyless manner, so that any
particular register $x$ appears in the update of at most one register.
The semantics of the operators on pairs is defined below:
$e_{1}\dadd e_{2}$ is defined to be  $\left(e_{1}.c\sradd e_{2}.c,e_{1}.d\right)$;
$e_{1}\dscale d$ equals $\left(e_{1}.c\srmul d,e_{1}.d\srmul d\right)$; and
$e_{1}\left[e_{2}\right]$ is given by $\left(e_{1}.c\sradd e_{1}.d\srmul e_{2}.c,e_{1}.d\srmul e_{2}.d\right)$.
While registers contain and expressions evaluate to pairs, the output function projects out
the ``$c$'' component of this pair: this is equivalent to instantiating the parameter $?$ to $0$, the additive
identity, since over semirings, the additive identity annihilates any other element under multiplication
($x.d \srmul 0 \sradd x.c = 0 \sradd x.c = x.c$).
The resulting model of CRA-definable cost functions is
$\CCF\left(\domain\times\domain,\dadd,\dscale d,\left[\cdot\right]\right)$

\begin{example}
\label{ex:4:1:3}Consider strings $w\in\left\{ a,b\right\} ^{*}$,
so that $f\left(w\right)$ is the number of $a$'s between the closest
pair of $b$'s. This function is in $\CF\left(\domain,\min,+c\right)$, but not in the more restricted classes:
$\CF\left(\domain,\min\left(\cdot,d\right),+c\right)$ and $\CF\left(\domain,+c\right)$. In figure \ref{fig:4:1:3}, we show
a $\CCF\left(\domain\times\domain,\dadd,\dscale,\left[\cdot\right]\right)$
machine that can compute $f$. The output in both $q_0$ and $q_1$ is identically $\infty$, while the output in $q_2$
is the ``$c$'' component of the ouput function $x$: $x.c$.

\begin{figure}
\centering{}\begin{tikzpicture}

  \node [state] (q0) {$q_{0}$};
  \node [coordinate,above left of=q0,node distance=7mm] (start) {};
  \node [state] (q1) [right=4cm of q0] {$q_{1}$};
  \node [state] (q2) [right=4cm of q1] {$q_{2}$};

  \draw [->] (start) to (q0);
  \draw [->] (q0) edge [loop above] node [smalltext]
    {$\begin{array}{c}
              a/ \\
              x := x \\
              y := y
     \end{array}$} (q0);
  \draw [->] (q0) edge node [smalltext, label=above:{$b/x:=x, y:=y$}] {} (q1);
  \draw [->] (q1) edge [smalltext, loop above] node
    {$\begin{array}{c}
              a/ \\
              x := x \\
              y := y \dscale 1
     \end{array}$} (q1);
  \draw [->] (q1) edge node [smalltext, text width=2cm, label=above:{$b/x:=y, \allowbreak y:=\left(\infty,0\right)$}] {} (q2);
  \draw [->] (q2) edge [smalltext, loop above] node
    {$\begin{array}{c}
              a/ \\
              x := x \\
              y := y \dscale 1
     \end{array}$} (q2);
  \draw [->] (q2) edge [smalltext, loop below] node
    {$\begin{array}{c}
              b/ \\
              x := x \dadd y \\
              y:=\left(\infty,0\right)
     \end{array}$} (q2);

 \node[below of=q1,node distance=15mm,smalltext]
      {$\begin{array}{l}
        \mu(q_0) = (\infty,\infty) \\
        \mu(q_1) = (\infty,\infty) \\
        \mu(q_2) = x
        \end{array}$};

\end{tikzpicture}\caption{\label{fig:4:1:3}The
$\CCF\left(\domain\times\domain,\dadd,\dscale,\left[\cdot\right]\right)$
machine for example \ref{ex:4:1:3}.}
\end{figure}

\end{example}

\mypar{Expressiveness}

The next theorem summarizes the relationship between functions definable by different CRA models. The rest of the session contains the proof of this theorem.

\begin{theorem}[Expressiveness of Semi-ring Cost Functions]\label{thm:semiring-exp}
If $\left(\domain,\sradd,\srmul\right)$ forms a semiring, then
\[
\CCF(\domain,\sradd,\srmul c)\ \subset\ \CCF\left(\domain\times\domain,\dadd,\dscale c,\left[\cdot\right]\right)\ =\
\reg{\domain,\sradd,\srmul d}\ \subset\ \CF(\domain,\sradd,\srmul c)
\]
\end{theorem}
We split the proof into the following lemmas.
\begin{lemma}
If $\left(\domain,\sradd,\srmul\right)$ forms a semiring, then
\[ \CCF(\domain,\sradd,\srmul c)\ \subset\ \CCF\left(\domain\times\domain,\dadd,\dscale c,\left[\cdot\right]\right) \]
\end{lemma}
\Proof~
We first show that the containment holds and then that it is strict.
Copyless CRAs with $\sradd$ and $\srmul c$ can be simulated by copyless CRAs operating over pairs and performing $\dadd$, $\dscale$, and $\left[\cdot\right]$. Given a $\CCF(\domain,\sradd,\srmul c)$ machine $M_{1}$, construct an $\CCF\left(\domain\times\domain,\dadd,\dscale c,\left[\cdot\right]\right)$ machine $M_{2}$ with the same states, and same registers. Replace every occurrence of $\sradd$ and $\srmul c$ in the update expressions to $\dadd$ and $\dscale c$ respectively.

To show strict containment, let $\left(\domain,\sradd,\srmul\right)$
be the tropical semiring. Our witness function is $f_{1}$ from Fig.
\ref{rcfex1}. First off, observe that $f_{1}\in\CCF\left(\domain\times\domain,\dadd,\dscale c,\left[\cdot\right]\right)$,
as shown in figure \ref{rcfex_f1}. We now demonstrate that $f_{1}\notin\CCF(\domain,\min,+c)$,
and our proof is similar to that of theorem \ref{thm:plus-c-copying-essential}.

\begin{figure}[t]
\centering
\begin{tikzpicture}[->,-stealth',semithick]
\node[state] (q0) {$q_0$};
\node[coordinate,node distance=7mm,above left of=q0] (start1) {};
\draw (start1) to (q0);

\draw (q0) to[loop left] node[smalltext, left] (t1) {} (q0);
\node[smalltext,node distance=5mm,left of=t1]
     {$\begin{array}{c}
              a/ \\
              x := x \dscale 1 \\
              y := y
     \end{array}$};

\draw (q0) to[loop above] node[smalltext, above] (t2)
        {$b\hspace{-.5em}\left/\hspace{-.8em}
         \begin{array}{l} x := x \\
                          y := y \dscale 1
         \end{array}\right.\hspace{-.5em}$}
      (q0);

\draw (q0) to[loop below] node[smalltext, below] (t3)
        {$e\hspace{-.5em}\left/\hspace{-.8em}
          \begin{array}{l} x := x \left[ y \right] \dscale 1 \\
                           y := \left( \infty, 0 \right)
          \end{array}\right.\hspace{-.5em}$}
      (q0);

\node[smalltext,node distance=19mm,below of=q0] {$\mu(q_0) = x$};


\end{tikzpicture}
\caption{$f_{1}$ from figure \ref{rcfex1} is in $\CCF\left(\domain\times\domain,\dadd,\dscale,\left[\cdot\right]\right)$. \label{rcfex_f1}}
\end{figure}

We proceed by contradiction. Say we are given a copyless CRA machine
$M$ over $\left(\domain,\min,+c\right)$ that implements $f$. Without
loss of generality, we can assume that in every update, a register
$x$ is either reset, or appears in its own update expression: $x:=\min\left(x+c,\ldots\right)$.

Consider a string of the form $w=b^{n}aeb^{n}ae\ldots$, where $n$
is greater than the number of states in $M$. Thus, for each $i$,
while processing the $i^{\mbox{th}}$ block of $b^{n}$, some state
$q_{i}$ (possibly depending on the block number $i$) must be visited
at least twice. For each $i$, let $x_{i}$ be the register which
influences the output after reading the input string $\left(b^{n}ae\right)^{i}a$.
Now observe that for all $i<j$, $x_{i}$ and $x_{j}$ are necessarily
distinct: if we pump $w$ to $w^{\prime}=\left(b^{n}ae\right)^{i-1}b^{n+k}ae\left(b^{n}ae\right)^{j-i}\ldots$,
the value of $x_{i}$ after $i$ blocks must be unchanged, but the
value of $x_{j}$ must change. We have thus established that a different
register must be used to produce the output after each block, but
this is not possible if the machine has only a finite number ($k$ in this case)
of registers.
\qed

\begin{lemma}
If $\left(\domain,\sradd,\srmul\right)$ forms a semiring, then
\[ \CCF\left(\domain\times\domain,\dadd,\dscale c,\left[\cdot\right]\right)\ =\ \reg{\domain,\sradd,\srmul d}. \]
\end{lemma}
\Proof~
From the definition of CRAs the terms constructed by the SSTTs are in correspondence with their most evaluated versions maintained by CRAs.
\qed

\begin{lemma}\label{lemma:reg-to-cra}
If $\left(\domain,\sradd,\srmul\right)$ forms a semiring, then
\[ \CCF\left(\domain\times\domain,\dadd,\dscale c,\left[\cdot\right]\right)\ \subseteq\ \CF(\domain,\sradd,\srmul c). \]
\end{lemma}
\Proof~
Consider a copyless CRA machine $M$ over $\left(\domain\times\domain,\dadd ,\dscale ,\left[\cdot\right]\right)$.
Let $V_{r}$ be the set of its registers. We construct a copyful CRA $M'$
over $\left(\domain,\sradd,\srmul c\right)$ equivalent to $M$.
We perform the following subset construction over registers. The states
and transitions of $M^{\prime}$ are the same as in $M$.
The set of registers $V_{l}$ of $M^{\prime}$ is the following:
\begin{enumerate}
\item $x.c$ and $x.d$ for every $x\in V_{r}$.
\item for every $S\subseteq V_{r}$, we maintain $d_{S}=\srmul_{x\in S}x.d$,
and for all $x\notin S$, $xd_{S}=x.c\srmul d_{S}$.
\end{enumerate}
The expression on the right of each of the above equalities is the
intended invariant we'll maintain. Because of the properties of the
semiring, we can simplify the resulting expression into a linear form (an expression of the form $\sradd_{i}\left(X_{i}\srmul a_{i}\right)\sradd c$, for some constants $a_i$ and $c$, where $i$ ranges over the registers).

We define an elementary update in $M$ as one in which:
the value of no register changes, or
exactly two registers permute: $\langle x,y\rangle:=\langle y,x\rangle$, or
exactly one register is reset: $x:=\left(c,d\right)$, or
exactly one register changes: $x:=x\dscale d$, or
exactly two registers change (addition): $\langle x,y\rangle:=\langle x\dadd y,\left(0,1\right)\rangle$, or
exactly two registers change (substitution): $\langle x,y\rangle :=\langle x\left[y\right],\left(0,1\right)\rangle$.
Observe that any copyless register update can be written as a finite
sequence of elementary updates. Also a finite sequence of updates
in a CRA machine over $\left(\domain,\sradd,\srmul c\right)$ can be summarized into a single update. Thus,
if we demonstrate a semantics-preserving transformation from elementary
updates to copyful linear updates, we are done.

Given a register $x\in V_{l}$, let $x$ be its value before the update,
and $x^{\prime}$ be its intended value after. We show that $x^{\prime}$
in each case can be written as a linear combination of the old values,
thus giving a linear update rule $x:=Expr$. Only the last two cases are interesting:

\begin{enumerate}
\item Addition: $\langle x,y\rangle:=\langle x\dadd y,\left(0,1\right)\rangle$.

\begin{enumerate}
\item $x.c^{\prime}=x.c\sradd y.c$, $x.d^{\prime}=x.d$. $y.c^{\prime}=0$
and $y.d^{\prime}=1$.
\item For $z\neq x,y$. $z.c^{\prime}=z.c$ and $z.d^{\prime}=z.d$.
\item For $S$, $x,y\notin S$. $d_{S}^{\prime}=d_{S}$, $xd_{S}^{\prime}=\left(x.c\sradd y.c\right)\srmul d_{S}^{\prime}=xd_{S}^{\prime}\sradd yd_{S}^{\prime}$.
$yd_{S}^{\prime}=0$, and $zd_{S}^{\prime}=zd_{S}$.
\item For $S$, $x\in S$, but $y\notin S$. $d_{S}^{\prime}=d_{S}$, $yd_{S}^{\prime}=0$,
and $zd_{S}^{\prime}=zd_{S}$. (Exactly the same as the previous case.)
\item For $S$, $x\notin S$, but $y\in S$. Let $S^{\prime}=S\setminus\left\{ y\right\} $.
$d_{S}^{\prime}=y.d^{\prime}\srmul d_{S^{\prime}}^{\prime}=d_{S^{\prime}}$.
$xd_{S}^{\prime}=x.c^{\prime}\srmul d_{S}^{\prime}=\left(x.c\sradd y.c\right)\srmul d_{S^{\prime}}=xd_{S^{\prime}}\sradd yd_{S^{\prime}}$.
$zd_{S}^{\prime}=z.c^{\prime}\srmul d_{S}^{\prime}=z.c\srmul d_{S^{\prime}}=zd_{S^{\prime}}$.
\item For $S$, $x,y\in S$. $S^{\prime}=S\setminus\left\{ x,y\right\} $.
$d_{S}^{\prime}=x.d^{\prime}\srmul y.d^{\prime}\srmul d_{S^{\prime}}^{\prime}=x.d\srmul d_{S^{\prime}}=d_{S^{\prime}\cup\left\{ x\right\} }$.
$zd_{S}^{\prime}=z.c^{\prime}\srmul d_{S}^{\prime}=zd_{S^{\prime}\cup\left\{ x\right\} }$.
\end{enumerate}
\item Substitution: $\langle x,y:=x\left[y\right],\left(0,1\right)\rangle$. This shows why we needed to keep the subset
registers.

\begin{enumerate}
\item $x.c^{\prime}=x.c\sradd x.d\srmul y.c=x.c\sradd yd_{\left\{ x\right\} }$.
$x.d^{\prime}=x.d\srmul y.d=d_{\left\{ x,y\right\} }$. $y.c^{\prime}=0$
and $y.d^{\prime}=1$.
\item For $z\neq x,y$, $z.c^{\prime}=z.c$ and $z.d^{\prime}=z.d$.
\item For $S$, $x,y\notin S$. $d_{S}^{\prime}=d_{S}$, $xd_{S}^{\prime}=x.c^{\prime}\srmul d_{S}^{\prime}=\left(x.c\sradd x.d\srmul y.c\right)\srmul d_{S}=xd_{S}\sradd yd_{S\cup\left\{ x\right\} }$.
$yd_{S}^{\prime}=0$, and $zd_{S}^{\prime}=zd_{S}$.
\item For $S$, $x\in S$ but $y\notin S$. Let $S^{\prime}=S\setminus\left\{ x\right\} $.
$d_{S}^{\prime}=x.d^{\prime}\srmul d_{S^{\prime}}^{\prime}=d_{\left\{ x,y\right\} }\srmul d_{S^{\prime}}=d_{S\cup\left\{ y\right\} }$.
$yd_{S}^{\prime}=0$, and $zd_{S}^{\prime}=z.c^{\prime}\srmul d_{S}^{\prime}=z.c\srmul d_{S\cup\left\{ y\right\} }=zd_{S\cup\left\{ y\right\} }$.
\item $x\notin S$, but $y\in S$. Let $S^{\prime}=S\setminus\left\{ y\right\} $.
$d_{S}^{\prime}=y.d^{\prime}\srmul d_{S^{\prime}}^{\prime}=d_{S^{\prime}}$.
$xd_{S}^{\prime}=x.c^{\prime}\srmul d_{S}^{\prime}=\left(x.c\sradd yd_{\left\{ x\right\} }\right)\srmul d_{S^{\prime}}=xd_{S^{\prime}}\sradd yd_{S^{\prime}\cup\left\{ x\right\} }$.
$zd_{S}^{\prime}=z.c^{\prime}\srmul d_{S}^{\prime}=z.c\srmul d_{S^{\prime}}=zd_{S^{\prime}}$.
\item Both $x,y\in S$. Let $S^{\prime}=S\setminus\left\{ x,y\right\} $.
$d_{S}^{\prime}=x.d^{\prime}\srmul y.d^{\prime}\srmul d_{S^{\prime}}^{\prime}=d_{\left\{ x,y\right\} }\srmul d_{S^{\prime}}=d_{S^{\prime}\cup\left\{ x,y\right\} }=d_{S}$.
$zd_{S}^{\prime}=z.c^{\prime}\srmul d_{S}^{\prime}=z.c\srmul d_{S}=zd_{S}$.\end{enumerate}
\end{enumerate}
\qed\\\\
Finally, the containment established by the above theorem is strict.

\begin{lemma}
Over the tropical semiring, there exist functions in $\CF\left(\domain,\sradd,\srmul c\right)$ which are not in $\reg{\domain,\sradd,\srmul d}$.
\end{lemma}
\Proof~
An example of such a function is $f_{3}$ in figure \ref{rcfex1}.
Regardless of the string $w$, $f\left(wb^{\left|w\right|}e\right)=\left|w\right|_{a}$.
Thus, in any state $q$, the machine has to contain, in some register
$x_{q}$, $\left|w\right|_{a}+c_{qx}$. However, for all $w$ and
$k>0$, there is some $\sigma$ so that $\left|f\left(w\sigma\right)-\left|w\sigma\right|_{a}\right|\geq k$.
Let's identify some witness for this by writing $\sigma\left(w,k\right)$.
In particular, this means that there has to be some register tracking
the value of the function, which is distinct from the register $x_{q}$
tracking $\left|w\right|_{a}$. We have thus established that at least
two registers are necessary.

Consider a machine with two registers. For the largest constant $c$
appearing in the description of the machine, consider the string $a^{c}\sigma\left(a^{c},c\right)$.
At this point, we have two registers - one containing the number of
$a$s, and the other containing the function. If we now feed the machine
a suffix $b^{\left|\sigma\left(\epsilon,c\right)\right|}c$, the value
of the function is equal to the number of $a$s in the input string.
The machine now has two choices: either copy the value, or choose
to track both the functions and the number of $a$s in the same register.
This latter choice cannot happen: for we can then feed the suffix
$\sigma\left(\sigma\left(\epsilon,c\right)b^{\left|\sigma\left(\epsilon,c\right)\right|}e\right)b^{\left|\sigma\left(\sigma\left(\epsilon,c\right)b^{\left|\sigma\left(\epsilon,c\right)\right|}e\right)\right|}e$,
and force the machine into making a mistake.

For multiple registers, we perform a multi-step pumping argument similar
to the above. Define the sequence: $\sigma_{1}=a^{c}\sigma\left(a^{c},c\right)$,
$\sigma_{s}=\sigma_{1}a^{c}b^{\left|\sigma_{1}a^{c}\right|}e$, \ldots{},
$\sigma_{i+1}=\sigma_{i}a^{c}b^{\left|\sigma_{i}a^{c}\right|}e$,
\ldots{} The argument involves observing after reading each $\sigma_{i}$,
the number of ``useful'' variables, in the absence of copyful assignments,
decreases by one: $\sigma_{1}\sigma_{2}\ldots\sigma_{i}\ldots$.
\qed

\mypar{Relation to Weighted Automata}
In Section~\ref{wa}, we noted that single-valued weighted automata correspond exactly to
CRAs with addition. Now we show that nondeterministic weighted automata and
(deterministic) CRAs (without the copyless restriction)
with $\sradd$ and $\srmul c$ express exactly the same class of functions.
The translation from weighted automata to CRAs can be viewed
as a generalization of the classical subset construction for determinization.

\begin{theorem}[Weighted Automata Expressiveness]\label{wa-cra}
If $\left(\domain,\sradd,\srmul\right)$ forms a semiring, then the
class of functions $\CF\left(\domain,\sradd,\srmul c\right)$
is exactly that representable by weighted automata.
\end{theorem}
\Proof~
Let $W=\left(\Sigma,P,I,F,E,\lambda,\rho\right)$ be a weighted
automaton. We construct the corresponding CRA $M$ over $\left(\domain,\sradd,\srmul\, c\right)$:
The set of states $Q=2^{P}$. The state set is obtained by the standard
subset construction. The intuition is that $M$ is in state $q\subseteq P$
after processing a string $w$ if $q$ is exactly the set of states
reached in $W$ after processing $w$. Thus, $\delta\left(q,a\right)=\left\{ p^{\prime}\in P\mid\exists p\in q,p\to_{W}^{a}p^{\prime}\right\} $.
The initial state $q_{0}$ is the set of initial states of the weighted
automaton, $I$.
The set of registers is $X=\left\{ x_{p}\mid p\in P\right\} $: there
is a register $x_{p}$ for every state $p\in P$. The following is the intuition behind these registers: consider some state $p$, and all paths from the set of initial states $I$ to $p$. Along each path, take the $\srmul$-product of the weights, and $\sradd$-add the values thus obtained for all paths. The intent is for $x_{p}$ to hold this value.
For each state $p\in I$, the register $x_{p}$ is initialized to
$\lambda\left(p\right)$. Even though in the definition of CRAs, registers were initialized to $0$ (or some other constant), by simply adding a new initial state which explicitly initializes registers before use, we can simulate registers being initialized to anything we choose.
When the CRA makes a transition $q\to^{a}q^{\prime}$, the register
update is given by $\forall p^{\prime}\in q^{\prime}$, $x_{p^{\prime}}:=\sradd\{x_{p}\srmul c\mid p\stackrel{a,c}{\rightarrow}_{W}p^{\prime}\}$.
That is, to obtain the value of $x_{p^{\prime}}$, we consider each
state $p$ such that there is an $a$-labeled transition from $p$
to $p^{\prime}$ with cost $c$ in the weighted automaton, add $c$
to $x_{p}$ according to $\srmul$, and take $\sradd$ over all such
values.
In state $q\subseteq P$, the output function is defined as $\sradd\left\{ x_{p}\srmul\rho\left(p\right)\mid p\in q\right\} $.
The output function is the $\sradd$-sum of all the product of all
paths.
To prove the correctness of this construction, observe the inductive
invariant: for all $w$, and for all states $p$, the register $x_{p}$
in the CRA $M$ contains the $\sradd$-sum of the $\srmul$-product
of the weights of all paths leading from some initial state to $p$
on $w$. This depends on the distributivity of $\srmul$ over $\sradd$.
Note that the same register $x_{p}$ contributes to all $x_{p^{\prime}}$s
for all its $a$-successor states $p^{\prime}$. Thus, the update
is not necessarily copyless.

In the reverse direction, let $M$ be a CRA over $\left(\domain,\sradd,\srmul c\right)$
with states $Q$ and registers $X$. Construct the following weighted
automaton $W$:
\begin{enumerate}
\item The state set $P$ is $Q\times X\cup Q$. Intuitively, a state $\left(q,x\right)\in Q\times X$
calculates transformations happening to individual registers, and
states $q\in Q$ calculates the constant offset possibly imposed by
the output function. Formally, after processing some word $w$, if
$M$ reaches state $q$, then the value of register $x$ is equal
to the value reaching state $\left(q,x\right)$: along each path from
some initial state to $\left(q,x\right)$, multiply all the weights,
and add the values thus obtained along all such paths. Also, such
paths exist iff processing $w$ takes $M$ to state $q$.
\item The initial states are $I=\left\{ q_{0}\right\} \times X\cup\left\{ q_{0}\right\} $.
All initial weights are equal to the multiplicative identity.
\item Say the output function at $q\in Q$ in $M$ is given by $\mu_{q}\left(\mathbf{x}\right)=\sradd_{i}\left(x_{i}\srmul a_{i}\right)\sradd c_{q}$.
Then, for each $i$, the output weight $\rho\left(q,x_{i}\right)=a_{i}$,
and $\rho\left(q\right)=c_{q}$.
\item For every transition $q\to^{a}q^{\prime}$ in $M$, create the transition
$q\to^{a,1}q^{\prime}$ in $W$. Also, for every variable update $x:=\sradd_{i}\left(x_{i}\srmul a_{i}\right)\sradd c_{x}$
that occurs during this transition, create the transitions $\left(q,x_{i}\right)\to^{a,a_{i}}\left(q^{\prime},x\right)$
in $W$ (for each $i$). Also add the transition $q\to^{a,c_{x}}\left(q^{\prime},x\right)$.
That the intended invariant is maintained follows from the distributivity
properties of a semiring.\end{enumerate}
\qed

\mypar{Decision Problems for Min-Plus Models}
Now we turn our attention to semirings in which the cost domain
is a numerical domain such as $\Nat\cup \left\{\infty\right\}$, $\sradd$ is the minimum operation, and $\srmul$ is the addition.
First let us consider shortest path problems for CRAs over the cost model $({\mathbb Q\cup \left\{\infty\right\}},\min,+c)$.
Given a CRA over such a cost model, we can construct a weighted automaton using the construction
in the proof of Theorem~\ref{wa-cra}.
Shortest paths in a weighted automaton can be solved in polynomial-time using standard algorithms \cite{mohri_weighted_2009}.

\begin{theorem}[Shortest Paths in CRAs over min and $+c$]
Given a CRA $\edwa$ over the cost model $(\Rat\cup \left\{\infty\right\},\min,+c)$, computing $\min\{\edwa(w)\sep w\in\fm\Sigma\}$
is solvable in $\ptime$.
\end{theorem}

It is known that the equivalence problem for weighted automata over the tropical semiring
is undecidable.
It follows that checking whether two CRAs over the cost model $(\Nat\cup \left\{\infty\right\},\min,+c)$ compute the same cost function,
is undecidable.
The existing proofs of the undecidability of equivalence rely
on the unrestricted non-deterministic nature of weighted automata, and thus on
the copyful nature of CRAs with $\min$ and $+c$.
We conjecture that the equivalence problem for copyless CRAs over $(\Nat\cup \left\{\infty\right\},\min,+c)$,
and also for the class $\reg{\Nat\cup \left\{\infty\right\},\min,+c}$ is decidable.

\section{Discounted Costs}\label{sec:disc}
In this section, we focus on the class of regular cost functions
definable using $+c$ and $*d$. Such cost functions allow both adding
costs and scaling by discount factors.

\mypar{Past Discounts} First let us focus on CRAs over the cost
model $\CostModel\left(\Rat, +c, *d\right)$. At every step, such a
machine can set a register $x$ to the value $d*x + c$: this
corresponds to discounting previously accumulated cost in $x$ by a
factor $d$, and paying an additional new cost $c$. We call such
machines the {\em past-discount} CRAs (see $f_4$ of
Figure~\ref{rcfex1} for an example). Note that the use of multiple
registers means that this class of cost functions is closed under
regular choice and regular look-ahead: the discount factors can
depend conditionally upon future events. It is easy to check that
the cost functions definable by past-discount CRAs belong to the
class $\reg{\Rat,+c,*d}$. Our main result for past-discount CRAs is
that the min-cost problem can be solved in polynomial-time. First,
multiple registers can be handled by considering a graph whose
vertices are pairs of the form $(q,x)$, where $q$ is a state of the
CRA, and $x$ is a register. Second, classical shortest path
algorithm can be easily modified when the update along an edge
scales the prior cost before adding a weight to it, this is
sometimes called {\em generalized shortest path\/}
(see~\cite{batagelj_generalized_2000,oldham_combinatorial_1999}).

\newcommand{\fp}{f_{pd}}
\newcommand{\ff}{f_{fd}}

\begin{theorem}[Shortest Paths for Past Discounts]
Given a past-discount CRA $\edwa$ over the cost model
$(\Rat,+c,*d)$, computing $\min\{\edwa(w)\sep w\in\fm\Sigma\}$ is
solvable in $\ptime$.
\end{theorem}
\Proof~
We reduce this problem to the {\em generalized shortest path\/}
 problem (see~\cite{batagelj_generalized_2000,oldham_combinatorial_1999}) on a graph, where edges are parameterized
 by cost $c(e)$ and weight $w(e)$ and the cost of a path $p=(e_1,\ldots,e_n)$ is
$c(p)=c(e_1)+w(e_1)*(c(e_2)+w(e_2)*(\ldots+w(e_{n-1})*c(e_{n})))$, while the weight of $p$ is $w(p)=w(e_1)*\ldots*w(e_n)$.

Consider a past-discount CRA $\edwa = (\inputalph, \edwastates, \edwainitst, \edwavariables, \edwatrans, \edwavarup, \edwafinal)$,
without loss of generality, assume that for each $\state,\state'\in \edwastates$, $a\in \inputalph$, if $\edwatrans(\state,a)=\state'$, then
for all $y\in \edwavariables$, $\edwavarup(\state,a,y)$ has the form $dx+c$ for some $x\in \edwavariables$; and for all $\state \in \edwastates$,
if $\edwafinal(\state)$ is defined, then it has the form $dx+c$ for some $x\in \edwavariables$.
We construct the following graph $G=(V,E)$. 
For each $\state \in \edwastates$ and $x \in \edwavariables$, $G$ has a vertex $(\state,x)$.
Moreover, $G$ has a source vertex $s$, a target vertex $t$. The graph $G$ maintains the invariant that
there is a path from $s$ to $(\state, x)$ if and only if there is a run of $M$ such that the value of $x$ in state $\state$ "flows" to the
final output.
\loris{This invariant is too informal, you should say something like:
The graph $G$ maintains the invariant that
there is a path $p$ from $s$ to $(\state, x)$
if and only if there exists a string $s\in\Sigma$
such that $M$ on $s$ reaches the configuration $(q,\valuation)$
and $\valuation(x)=c(p)$ (or something like that).
}
Formally, for each $\state,\state' \in \edwastates$ and $a\in \inputalph$, such that
$\edwatrans(\state,a)=\state'$, for any register $y\in \edwavariables$:
if $\edwavarup(\state,a,y)=dx+c$, for some $x$, $G$ has an
edge $e$ from $(\state',y)$ to $(\state,x)$ with 
weight $w(e)=d$ and  cost $c(e)=c$.

Finally,
for all $\state\in \edwastates$, if $\edwafinal(\state)=dx+c$, $G$ has
an edge $e$ from $s$ to $(\state,x)$ with 
weight $w(e)=d$, cost $c(e)=c$;
for each $x\in \edwavariables$, $G$ has an edge $e$ from $(\edwainitst,x)$ to $t$,  with 
weight $w(e)=0$, cost $c(e)=0$.

Given a run $(\edwastate_0,\valuation_0)\ldots
(\edwastate_n,\valuation_n)$ of $\edwa$ on input string $w=w_1\ldots w_n$, if the output is defined,
we claim that there is a $s-t$ path $p$ in $G$ such that the cost of $p$ is equal to the output of $\edwa$ on $w$.
For convenience, let's define $r_i$ be the register that contributes to final output on state $\state_i$,
i.e. $\edwafinal(\state_n)=dr_n+c$; $\edwavarup(\state_i,w_i,r_{i+1})=dr_i+c$ for all $i<n$.
Let's define $f_i(r_i)$ be the function that outputs on state $\state_i$, i.e. $f_n(r_n)=\edwafinal(\state_n)$,
and for $i<n$, $f_i(r_i)=f_{i+1}(\edwavarup(\state_i,w_i,r_{i+1}))$.
It's clear that $f_0(0)$ is the final output of $\edwa$ on $w$.
Let $p_i$ be the path $(s,v_n,\ldots,v_i)$, where $v_j=(\state_j,r_j)$. It is easy to see that the cost of $p_0$
is equal to the cost of the path $p=(s,v_n,\ldots,v_0,t)$.
We inductively prove that $f_i(r_i)=w(p_i)r_i+c(p_i)$.
In the base case, if $\edwafinal(\edwastate_n)=dr_n+c$, by construction, $c(s,v_n)=c$ and $w(s,v_n)=d$.
Therefore, $f_n(r_n)=dr_n+c=w(p_n)r_n+c(p_n)$.
Suppose, for all $i\ge k$, the inductive invariant holds.
If $\edwavarup(\state_{k-1},w_k,r_k)=dr_{k-1}+c$, then
$f_{k-1}(r_{k-1})=f_k(dr_{k-1}+c)=w(p_k)(dr_{k-1}+c)+c(p_k)=dw(p_{k})r_{k-1}+w(p_k)c+c(p_k)=w(p_{k-1})r_{k-1}+c(p_{k-1})$, since $w(v_k,v_{k-1})=d$ and $c(v_k,v_{k-1})=c$.
Therefore, the output of $\edwa$ on $w$ $f_0(0)=c(p_0)=c(p)$.
On the other hand, given an $s-t$ path $p=(s,(\state_n,x_n),\ldots,(\state_0,x_0),t)$,  it is easy to see that there exists a run  $(\edwastate_0,\valuation_0)\ldots
(\edwastate_n,\valuation_n)$ of $\edwa$ on some string $w$, that the output of $\edwa$ is equal to $c(p)$.
\qed

\mypar{Future Discounts} Symmetric to past discounts are future
discounts: at every step, the machine wants to pay an additional new
cost $c$, and discount all future costs by a factor $d$. While
processing an input $w_1\ldots w_n$, if the sequence of local costs
is $c_1,\ldots c_n$ and discount factors is $d_1,\ldots d_n$, then
the cost of the string is the value of the term $(c_1 + d_1*(c_2 +
d_2*(\cdots )))$. {\em Future-discount} CRAs are able to compute
such cost functions using registers that range over $\Rat\times\Rat$
and substitution: each register holds a value of the form $(c,d)$
where $c$ is the accumulated cost and $d$ is the accumulated
discount factor, and updates are defined by the grammar $
e:=(c,d)|e[c,d]|x$. The interpretation for $e[c,d]$ is defined to be
$(e.c+c*e.d,e.d*d)$ (that is, the current discount factor $e.d$ is
scaled by new discount $d$, and current cost $e.c$ is updated by
adding new cost $c$, scaled by the current discount factor $e.d$).
Like past-discount CRAs,future-discount CRAs are closed under
regular choice and regular look-ahead. Processing of future
discounts in forward direction needs maintaining a pair consisting
of cost and discount, and the accumulated costs along different
paths is not totally ordered due to these two objectives. However,
if we consider paths in ``reverse'', a single cost value updated
using assignments of the form $x:=d*x + c$ as in past-discount CRAs
suffices.

\begin{theorem}[Shortest Paths for Future Discounts]
Given a future-discounted CRA $\edwa$ over the cost model
$(\Rat,+c,*d)$, computing $\min\{\edwa(w)\sep w\in\fm\Sigma\}$ is
solvable in $\ptime$.
\end{theorem}
\Proof~
We reduce this problem to the {\em generalized shortest path\/} problem
(see~\cite{batagelj_generalized_2000,oldham_combinatorial_1999}) on a graph.

Consider a future-discount CRA $\edwa = (\inputalph, \edwastates, \edwainitst, \edwavariables, \edwatrans, \edwavarup, \edwafinal)$.
First we construct an equivalent future-discount CRA $\edwa' $ that every updating function and output function has the form $x[c,d]$. Formally,
$\edwa' = (\inputalph, \edwastates, \edwainitst, \edwavariables', \edwatrans, \edwavarup', \edwafinal')$, where
 $\edwavariables' = \edwavariables \cup \{\varepsilon\}$, such that $\varepsilon \not \in \edwavariables$. For each $\state,\state'\in \edwastates$,
 and $a\in \inputalph$, such that $\edwatrans(\state,a)=\state'$, for each $y\in \edwavariables$:
\begin{enumerate}
\item if $\edwavarup(\state,a,y)=x[c,d]$ for some $x$, $\edwavarup'(\state,a,y)=x[c,d]$;
\item if $\edwavarup(\state,a,y)=(c,d)$, $\edwavarup'(\state,a,y)=\varepsilon[c,d]$;
\item $\edwavarup'(\state,a,\varepsilon)=\varepsilon(0,1)$.
\end{enumerate}
For each $\state \in \edwastates$, $\edwafinal'(\state)=\varepsilon(c,d)$ if $\edwafinal(\state)=(c,d)$,
$\edwafinal'(\state)=\edwafinal(\state)$ otherwise.
Second, we construct the graph $G=(V,E)$ with source $s$ and target $t$, such that there is a path $p$ from $s$ to $v$ if and only if
there is a run $(\edwastate_0,\valuation_0)\ldots
(\edwastate_n,\valuation_n)$ of $\edwa'$ and some $x\in \edwavariables$, such that $\valuation_n(x)=(c(p),w(p))$.
Formally, for each $\state \in \edwastates$, and each $x \in \edwavariables'$, $G$ has a vertex $(\state,x)$.
Moreover, $G$ has a source vertex $s$, a target vertex $t$. For each
$\state,\state'\in \edwastates$ and $a\in \inputalph$, such that
$\edwatrans(\state,a)=\state'$,  for each register $y\in \edwavariables$:
if $\edwavarup'(\state,a,y)=x[c,d]$ for some $x$, $G$ has an
edge $e$ from $(\state,x)$ to $(\state',y)$, with weight $w(e)=d$, cost $c(e)=c$;
Finally,
for each $\state\in \edwastates$, if $\edwafinal'(\state)=x[c,d]$, $G$ has
an edge $e$ from $(\state,x)$ to $t$ with weight $w(e)=d$, cost $c(e)=c$;
for each $x\in \edwavariables$, $G$ has an edge $e$ from $s$ to $(\edwainitst,x)$, with weight $w(e)=1$, cost $c(e)=0$.

Given a run $(\edwastate_0,\valuation_0)\ldots
(\edwastate_n,\valuation_n)$ of $\edwa'$ on input string $w=w_1\ldots w_n$, if the output is defined,
we claim that there is a $s-t$ path $p$ in $G$ such that the cost of $p$ is equal to the output of $\edwa'$ on $w$.
Let's define $r_i$ be the register that contributes to final output on state $\state_i$,
i.e. $\edwafinal'(\state_n)=r_n[c,d]$; $\edwavarup'(\state_i,w_i,r_{i+1})=r_i[c,d]$ for all $i<n$.
Let $p_i$ be the path $(s,v_0,\ldots,v_i)$, where $v_j=(\state_j,r_j)$ and $p=(s,v_0,\ldots,v_n,t)$.
For simplicity. we abuse the notation of the name of register and functions to mean their values under
interpretation.
We inductively prove that $r_i=(c(p_i),w(p_i))$.
In the base case, $r_0=(0,1)=(c(p_0),w(p_0))$ by construction.
Suppose, for all $i\le k$, the inductive invariant holds.
If $\edwavarup(\state_{k},w_k,r_{k+1})=r_{k}[c,d]$, then
$r_{k+1}=(r_k.c+c*r_k.d,r_k.d*d)=(c(p_k)+c*w(p_k),w(p_k)*d)=(c(p_{k+1}),w(p_{k+1}))$,since by construction,
$c(v_k,v_{k+1})=c$ and $w(v_k,v_{k+1})=d$.
Therefore, the output of $\edwa'$ on $w$ $\edwafinal'(\state_n)=r_n[c,d]=(r_n.c+c*r_n.d,r_n.d*d)=
(c(p_n)+c*w(p_n),w(p_n)*d)=(c(p),w(p))$, since by construction $c(v_n,t)=c$ and $w(v_n,t)=d$.
On the other hand, given an $s-t$ path $p=(s,(\state_0,x_0),\ldots,(\state_n,x_n),t)$,  it is easy to see that there exists a run  $(\edwastate_0,\valuation_0)\ldots
(\edwastate_n,\valuation_n)$ of $\edwa'$ on some string $w$, that the output of $\edwa'$ is equal to $c(p)$.
\qed

\mypar{Global Discounts} A {\em global-discount CRA\/} is capable of
scaling the global cost (the cost of the entire path) by a discount
factor. As in case of future-discount CRAs, it uses registers that
hold cost-discount pairs. We now assume that discounts range over
$[0,1]$ and costs range over $\posrat$. The registers are updated
using the grammar $e:=(0,1)\sep e \gdplus (c,d) \sep x$. The
interpretation for $e\gdplus (c,d)$ is defined to be $(d*e.c+ e.d*c,
e.d*d)$ (that is, the current discount factor $e.d$ is scaled by new
discount $d$, and current cost $e.c$ is updated by first scaling it
by the new discount, and then adding new cost $c$ scaled by the
current discount factor $e.d$). Analyzing paths in a global-discount
CRA requires keeping track of both the accumulated cost and
discount. We can show a pseudo-polynomial upper bound; it remains
open whether there is a strongly polynomial algorithm for shortest
paths for this model:

\begin{theorem}[Shortest Paths for Global Discounts]
Given a global-discount CRA $\edwa$ over the cost model
$(\posrat,+c,[0,1],*d)$ and a constant $K\in\posrat$, deciding
$\min\{\edwa(w)\sep w\in\fm\Sigma\}\le K$ is solvable in \np.
Computing the minimum is solvable in \ptime\ assuming increments are
restricted to adding natural numbers in unary encoding.
\end{theorem}
\Proof~
Consider a global-discount CRA $\edwa = (\inputalph, \edwastates, \edwainitst, \edwavariables, \edwatrans, \edwavarup, \edwafinal)$
, we construct a graph $G=(V,E)$, that each edge $e\in E$ is parameterized with a cost $c(e)$ and a discount $d(e))$ as we did above.
For a path $p=(e_1,\ldots,e_n)$,
the cost of the path is defined as $c(p)=(c(e_1),d(e_1))\gdplus (c(e_2),d(e_2))\gdplus \ldots$ $\gdplus (c(e_n),d(e_n))$ $=\sum_i c(e_i) \prod_i d(e_i)$.
Following the proof above, it is easy to see solving the shortest path in $\edwa$ is equivalent to solving the shortest path in $G$.

To prove the \np bound, we first observe that if there is a reachable cycle that
contains an edge $e$ with $d(e)<1$, then repeating this cycle
drives the global discount to 0, and thus, existence of such a cycle implies
that the min-cost (at the limit) is 0.
Notice that the shortest path doesn't need to involve a cycle
in which all discount factors are equal to 1 (since costs are non-negative).
The \np-bound follows from following fact. We can write an \np algorithm that guesses: 1) a reachable cycle and verifies if there is an edge $e$ with $d(e)<1$ in this cycle, or 2) a simple path and verify if its cost is less than $K$.
Suppose incremental costs $c_i$'s are small natural numbers.
The pseudo-polynomial algorithm for this case relies on the following idea: for a given value $c$ and a vertex $v$,
computing the ``best'' global discount over all paths from source to $v$ with
sum of incremental costs equal to $c$, can be solved by adopting shortest path algorithms, and
the set of interesting choices of $c$ can be bound by $nb$ for a graph with $n$ vertices if
each increment is a number between $0$ to $b$. Thus a variation of Bellman-Ford suffices (See algorithm \ref{shortestpath}).
\begin{algorithm}
\caption{Shortest Path Algorithm on Global Discount CRA }
\label{shortestpath}
\begin{algorithmic}
    \STATE \slash \slash $Discount[i,v,c]$ stores the best global discount from $s$ to $v$ among paths with length $\le i$, when the sum of incremental cost is $c$.
    \STATE $Discount[0,v,c]=\infty$, for every node $v$ and $0< c\leq nb$
    \STATE $Discount[0,s,0]=1$
    \FOR {$i=1$ to $n$}
        \FORALL {$v$ in $G$ and $0<c\leq nb$}
            \STATE $Discount[i,v,c]=\min\{Discount[i-1,v,c],\min_{e=(u,v)}\{Discount[i-1,u,c-c(e)]*d(e)\}\}$
        \ENDFOR
    \ENDFOR
    \STATE \slash \slash checks if there is a cycle with discount $\le 1$
    \IF {$\exists v,c$, s.t. $Discount[n-1,v,c]\not = Discount[n,v,c]$}
        \STATE return $0$
    \ENDIF
    \STATE \slash \slash Output the best simple path
    \RETURN $\min_{0<c\leq nb}\{Discount[n-1,t,c]*c\}$
\end{algorithmic}
\end{algorithm}
\qed

\mypar{Regular Functions for Inc-Scale Model} The class of regular
functions for the cost model $(\Rat,+c,*d)$ is defined via
SSTTs over the inc-scale grammar $\CG(+c,*d)$. It is to show that:

\begin{theorem}[Expressiveness of Inc-Scale Models]
The cost functions definable by past-discount CRAs, by
future-discount CRAs, and by global-discount CRAs all belong to
$\reg{\Rat,+c,*d}$.
\end{theorem}

The min-cost problem for this class of functions is still open.
However, we can show the equivalence problem to be decidable. First,
using the construction similar to the one used to establish
$\reg{\domain,\mytimes,\myplus c}\ \subset\
\CF(\domain,\mytimes,\myplus c)$ (see
Theorem~\ref{thm:semiring-exp}), we can represent cost functions in
$\reg{\Rat,+c,*d}$ using (copyful) CRAs that use $+$ and $*d$. Such
CRAs have {\em linear\/} updates, and the algorithm for checking
equivalence of CRAs with addition can be used for this case also.
\begin{theorem}
Given two functions $f_1,f_2\in \reg{\Rat,+c,*d}$ represented by SSTTs
over the cost grammar $\CG(+c,*d)$, checking whether the two
functions coincide, can be solved in time polynomial in the number
of states and exponential in the number of registers.
\end{theorem}
\Proof~
Given an \SSTT $T$ for $f \in \reg{\domain,+c,*d}$, we use lemma \ref{lemma:reg-to-cra} to construct an equivalent
CRA $\edwa$ using $+$ and $*d$. $\edwa$ has the same number of states as $T$, and an exponential number of variables.
We claim that checking equivalence between two CRAs with $+$ and $*d$ is solvable in time polynomial in the number of
states and number of variables (the proof is similar to that of theorem \ref{thm:equiv-CRA-add}). Thus, checking the equivalence
of two functions $f_1$ and $f_2$ expressed as STTs over $\CG(+c, *d)$ can be done in time polynomial in the number of
states and exponential in the number of variables. \qed

\section{Related Work}\label{sec:rel}
\paragraph{Weighted Automata (\WA) and Logics.}
Finite-state \WA have been an active area of research, with numerous
articles studying their algebraic and algorithmic properties. See
\cite{droste_handbook_2009} for a comprehensive exposition.  An important
problem for \WA is that of determinization \cite{mohri_weighted_2009,
kirsten_determinization_2005}. A deterministic \WA is defined in the usual
sense: no two outgoing transitions from a state share the same input
label.  It has been shown that there are \WA that do not admit equivalent
deterministic \WA. In contrast, the cost register automata that we
introduce in this paper are deterministic machines with equivalent
expressive power and equally efficient decision problems as weighted
automata. We believe that this makes them a more suitable model for
expressing weighted computations.

It has been shown that the equivalence problem for \WA over the tropical
semiring is undecidable using a reduction from Hilbert's tenth problem
\cite{krob_equality_1992}, and by a reduction from the halting problem for
two counter machines \cite{almagor_what_2011}. The only known class of
weighted automata over the tropical semiring with decidable equivalence problem is that of
finite-valued weighted automata \cite{weber_finite_1994}. For a given $k$,
a weighted automaton is said to be $k$-valued if the number of distinct
values computed along all accepting paths is at most $k$. A weighted
automaton is called finite-valued if there exists a $k$ such that it is
$k$-valued.  We conjecture that the equivalence problem for \EDWA over
$min$ and $+c$ with the copyless restriction is decidable. If this is
true, it would give the largest known class with decidable equivalence.
In \cite{Kiefer} the authors provide a randomized algorithm
to solve equivalence of weighted automata over the semiring with addition and
scaling.

In \cite{droste_weighted_2005}, the authors discuss a weighted
MSO-logic that disallows universal second order quantification and
places restrictions on universal first order quantification. The
authors show that the formal power series definable in this logic
coincides with the set of behaviors of weighted automata. In contrast,
in this paper, we introduce automata and machines that exactly capture
MSO-definable cost functions.

\paragraph{Discounted weighted computations and Generalized Shortest Paths.}
Generalized network flow problems extend flow problems on directed
graphs by specifying multipliers on edges in addition to costs
\cite{goldberg_combinatorial_1988, oldham_combinatorial_1999}.  The
problem of finding the minimum cost flow (which in some cases is
equivalent to the shortest distance path) from a source to a target
can be solved in polynomial time \cite{oldham_combinatorial_1999,
batagelj_generalized_2000}.  Future discount machines that we
introduce in this paper provide a nice formalism that subsumes such
problems, and have strongly polynomial time algorithms for determining
the minimum cost path.  In this paper, we also introduce past discount
and global discount machines that also have efficient algorithms for
determining the minimum cost paths.

In \cite{droste_weighted_2009}, the authors introduce weighted logic
for infinite words. In order to address convergence of the weighted
sum, the authors assume discounting along later edges in a path (\ie,
future discounting). Extending the results of this paper to discounted
weighted computations over infinite words remains open.

\paragraph{Transducer Models.}
A wide variety of different models have been proposed to model string
and tree transductions. The models that are most relevant to this
paper are MSO-definable transductions
\cite{courcelle_graph_2002,engelfriet_mso_2001} and macro tree
transducers \cite{engelfriet_macro_1985,engelfriet_macro2_1999}. An
MSO-definable graph transduction specifies a function between sets of
graphs; the nodes, edges and labels of the output graph are described
in terms of MSO formulas over the nodes, edges and labels of a finite
number of copies of the input graph.  A macro tree transducer (MTT) is
a top-down tree to tree transducer equipped with parameters.
Parameters can store temporary trees and append them to the final tree
during the computation. In general, MTT are more expressive than
MSO-definable tree transductions. A subclass of MTTs obtained by
restricting the number of times a subtree and a parameter can be used
has been shown to be equi-expressive as MSO-definable tree
transductions \cite{engelfriet_macro2_1999}. In addition to these
models, formalisms such as attribute grammars
\cite{engelfriet_macro2_1999}, attribute tree transducers
\cite{bloem_comparison_2000} have also been studied.

Streaming tree transducers \cite{alur_stt_2011} (STTs), introduced by
two of the co-authors in this paper are a new formalism for expressing
MSO-definable tree-to-tree and string-to-tree transductions. In
comparison to some of the transducer models discussed above, STTs have
distinguishing features that make them desirable as a canonical model
for specifying regular or MSO-definable transductions: (1) STTs
produce the output in linear time by performing a single pass over the
input, (2) they preserve desirable properties such as closure under
sequential composition and regular look-ahead, and (3) they have good
algorithmic properties such as decidability of functional equivalence.

\paragraph{Regularity over Data Languages.} Data languages allow
finite strings over data values that can be drawn from a possibly
infinite data domain \cite{neven_finite_2004},
\cite{kaminski_finite_1994} \cite{bjorklund_notions_2010}. Register
automata are often used as acceptors for data languages. A key feature
of such automata is that they allow registers to store and test data
values.  Beyond the similarity in nomenclature, register automata that
are studied in this line of work are quite distinct from cost register
automata introduced in this paper. The former are essentially defined
over an infinite input alphabet, and the critical difference lies in
the fact that almost every variant of data automata allows testing
equality of data values, which mostly causes interesting decision
problems to become undecidable. Cost register automata use the cost
registers in a strictly write-only fashion, which makes them
incomparable to variants of data automata that use read/write
registers.


\paragraph{Regular Cost Functions.}
In \cite{colcombet_theory_2009}, Colcombet defines a regular cost
function as a mapping from words to $\mathbb{N}^\omega$ (the set of
nonnegative integers and the ordinal $\omega$). A cost function is
precisely defined as an equivalence class over mappings from the set
of words to $\mathbb{N}^\omega$, such that functions $f$ and $g$ are
in the same equivalence class if for all words $w$, $f(w)$ is bounded
by some constant iff $g(w)$ is bounded is bounded by some constant.
The author then defines two classes of automata ($B$- and
$S$-automata), each of which uses a finite set of counters and allows
the counters to be incremented, reset or checked for equality with a
constant. The operational semantics of these automata are that the
automaton computes the least upper bound or the greatest lower bound
over the set of counter values encountered during its run. A cost
function is then called regular if it is accepted by a
history-deterministic $B$- or $S$-automaton.  The author also provides
an algebraic characterization of regular cost functions in terms of
stabilization monoids and equates recognizability of cost functions
with regularity. In \cite{colcombet_regular_2010}, the authors extend
this notion to regular cost functions over trees.

It is clear that the notions proposed in this line of work are
orthogonal to our characterization of regularity of cost functions.
The authors state that the motivation for the work in
\cite{colcombet_theory_2009,colcombet_regular_2010} is preserving nice
algorithmic and closure properties of regular languages for problems
such as equivalence and projection. However, the integer values in
these functions are considered modulo an equivalence which preserves
existence of bounds on the function values, but not the values
themselves. We believe that the notion of regularity of cost functions
that we propose in this paper is closer to the classical notions of
regularity such as MSO-definability.

\paragraph{Affine Programs.} In \cite{karr_affine_1976}, and more
recently in \cite{olm_note_2004, olm_precise_2004}, the authors
present the problem of deriving affine relations among variables of a
program.  An affine relation is a property of the form $a_0 +
\sum^{n}_{i=1} a_i.\mathbf{v_i} = 0$, where
$\mathbf{v_1},\ldots,\mathbf{v_n}$ are program variables that range
over a field such as the rationals or reals and  $a_i$ are constants
over the same domain. An affine program is a program with
nondeterministic branching where each edge of the program is labeled
with an assignment statement of the form $v_1 := v_2 + 2.v_3 + 3$,
\ie, where the RHS is an affine expression. We could define a \EDWA
over the cost model $\CostModel(\Rat, +, \Rat, *d)$, with the cost
grammar $t:= +(t,t) \sep *(t,d) \sep c$. An affine program is then
simply obtained by ignoring the input labels of the transitions in
such a \EDWA.  While the cost functions defined by such \EDWA do not
have interesting regularity properties, we remark that the equivalence
of such \EDWA can be checked in polynomial time by using ideas similar
to the ones in \cite{olm_note_2004}.

\paragraph{Quantitative Languages.} A quantitative language
\cite{CDH10,almagor_what_2011,AKL10} over infinite words is a function
$\Sigma^\omega \mapsto \mathbb{R}$. Such languages are generated by
weighted automata, where the value of a word $w$ is set as the maximal
value of all runs over $w$. By defining various value functions such
as $\mathit{Max}$, $\mathit{Sum}$, $\mathit{LimSup}$,
$\mathit{LimInf}$, different values can be computed for the run of a
weighted automaton on the string $w$.  Quantitative languages use the
fixed syntax of weighted automata, and thereby restricted to having a
single weight along each transition in their underlying automata.
Moreover, they face similar difficulties in determinization: for
interesting models of value functions, the corresponding automata
cannot be determinized.  An extension of \EDWA\ to $\omega$-regular
cost functions could prove to be a more expressive and robust model to
specify quantitative languages and to analyze their decision problems.

\section{Conclusions}\label{sec:conc}
\usetikzlibrary{patterns}

\begin{figure}
\subfloat[Hierarchy for CRA models]{
\label{fig:hierarchy}
\begin{minipage}[t]{0.49\textwidth}
    \begin{tikzpicture}[fill opacity=0.2]
        \tikzstyle{textz}=[font=\fontsize{8}{8}\selectfont,text opacity=1,text=black]

        \draw[thick] (0.9,0.7) rectangle (4.9,3.0);
        \node at (0.9,1.0) [textz,right] {Global Discounts};

        \draw[very thick] (-0.2,0.5) rectangle (7.3,3.0);
        \node at (4.85,1.0) [textz, right] {Reg $(\domain,\mytimes c,\myplus d)$};
        \node at (4.85,0.7) [textz, right] {$\equiv$ CRA Inc-Scale};

        \draw (0.9,1.2) rectangle (4.9,3.0);
        \node at (0.9,2.8) [textz,right] {Reg $(\domain, \mytimes c) \equiv$};
        \node at (0.9,2.45) [textz,right] {Reg $(\domain, \mytimes) \equiv$};
        \node at (0.9,2.1) [textz,right] {CRA $(\mytimes c) \equiv$};
        \node at (0.9,1.75) [textz,right] {CopylessCRA $(\mytimes) \equiv$};
        \node at (0.9,1.4) [textz,right] {Single-valued \WA};

        \draw[thick] (-0.1,1.2) rectangle (4.9,3.0);
        \node at (-0.1,2.8) [textz,right] {Past};
        \node at (-0.1,2.5) [textz,right] {Disc-};
        \node at (-0.1,2.2) [textz,right] {ounts};

        \draw[thick] (0.9,1.2) rectangle (6.3,3.0);
        \node at (4.85,2.8) [textz,right] {Future};
        \node at (4.85,2.5) [textz,right] {Discounts};

        \draw[semithick] (0.9,1.2) rectangle (4.9,3.4);
        \node at (0.85,3.2) [textz,right] {CopylessCRA $(\mytimes,\myplus c)$};

        \draw[thick] (0.9,1.2) rectangle (4.9,3.8);
        \node at (0.9,3.6) [textz,right] {Reg $(\domain, \mytimes, \myplus c)$};

        \draw[ultra thick] (0.9,1.2) rectangle (4.9,4.6);
        \node at (0.9,4.35) [textz,right] {CRA $(\mytimes,\myplus c) \equiv$};
        \node at (0.9,4.0) [textz,right] {Weighted Automata};

    \end{tikzpicture}
\end{minipage}
} \hfil
\subfloat[Complexity of Decision Problems]{
    \label{fig:complexity}
\begin{minipage}[t]{0.49\textwidth}
{\small
\begin{tabular}[b]{||l|c|c||}
\hline\hline
CRA with              &  Equivalence            & Min-Cost \bigstrut \\
\hline\hline
$(+c)$                &  \ptime                 & \ptime    \\
\hline
Copyless $(+)$       &  \ptime                 & \exptime   \\
\hline
$(min,+c)$            &  Undecidable            & \ptime     \\
\hline
Copyless $(min,+c)$  &  ?                      & \ptime     \\
\hline
Past-discounts        &                         & \ptime     \\
\cline{1-1} \cline{3-3}
Future-discounts      &  Poly in states      & \ptime     \\
\cline{1-1} \cline{3-3}
Global-discounts      &  Exp in registers & Pseudo-Poly \\
\cline{1-1} \cline{3-3}
Inc-Scale             &                         & ? \\
\hline\hline
\end{tabular}}
\end{minipage}
}
\caption{Summary of Results\label{results}}
\end{figure}

We have proposed a new approach to define regular functions for associating costs with strings.
The results for various classes of functions are summarized in Figure~2.
We hope that our work provides new insights into the well-studied topic of weighted automata, and
opens a whole range of new problems.
First, it is plausible that there is a compelling notion of congruences and canonicity for CRAs with increment.
Second, the decidability of copyless-CRAs with $\min$ and increment remains an intriguing open problem.
Third, we don't have algorithms for the min-cost problem for the class of regular functions with increment and scaling.
While we have not succeeded even in establishing decidability, we suspect that this problem admits efficient
approximation algorithms.
Fourth, we have considered only a small set of combinations of operations; studying the effects of adding operators
such as {\it max} would be worthwhile. Fifth, our notion of regularity and cost register automata
for mapping strings to costs can be extended to infinite strings and trees, as well as to timed and probabilistic systems.
Finally, we would like to explore practical applications: our framework seems suitable for expressing
complex, yet analyzable, pricing policies, say, for power distribution.

\newcommand{\BIBdecl}{\setlength{\itemsep}{0em}}
\bibliographystyle{abbrv}
\bibliography{../rcf}

\end{document}